\documentclass[%
 reprint,
 amsmath,amssymb,
 aps,
 amsfonts,
prd,
]{revtex4-2}
\usepackage[colorlinks=true,linkcolor=black,citecolor=blue,urlcolor=black]{hyperref}
\usepackage[mathlines]{lineno}
\usepackage{graphicx}
\usepackage{dcolumn}
\usepackage{bm}
\usepackage{latexsym}
\usepackage{booktabs}

\usepackage{csquotes}
\usepackage{subcaption}
\begin{document}

\preprint{APS/123-QED}

\title{Quasinormal modes of a charged covariant loop quantum black hole}

\author{Biao Zhou}%
 \email{Corresponding Author; biaozhou@hgnu.edu.cn}
\affiliation{%
School of Physics and Telecommunications,  Huanggang Normal University, Huanggang  438000, China
}%

\begin{abstract}
This paper focuses on the quasinormal modes of a charged loop quantum black hole that preserves diffeomorphism covariance. We have examined the influence of the black hole charge $Q$ and the quantum parameter $\zeta$ on the quasinormal modes of neutral massless scalar field, neutral massive scalar field and massless charged scalar field. For the neutral massless scalar field, our work shows that as in the case of the corresponding uncharged black hole that has been examined by other authors, the turn-on of $\zeta$ renders the real parts of the overtones a
tendency to approach zero. Increasing $Q$ can inhibit this tendency. For the neutral massive scalar field, when $Q$ and $\zeta$ are turned on, arbitrarily long-lived modes still exist. We also observe a non-monotonic behavior in the fundamental mode as the field mass varies for the first time. For the massless charged scalar field, our study shows that the asymptotic behaviors of the quasinormal modes of charged field and uncharged field are different as the extremal charge $Q$ is approached. We also find that the real part of the fundamental mode tends to zero at some critical value $eQ$ for a charged scalar field with field charge $e$.
\end{abstract}

\maketitle


\section{Introduction}
Einstein's theory of gravitation, general relativity, has passed various observational tests\cite{will_confrontation_2014}. However, there are still motivations to modify the theory. One of the motivations is to resolve the issue of singularity. Spacetime singularities are present in most physically relevant solutions of the Einstein equations. Spacetime curvature diverges at the singularity and classical physical theory fails there\cite{1965PhRvL..14...57P,1968PhRvL..20..878J,1966RSPSA.294..511H,Joshi2011}. Various schemes have been proposed to address the issue of singularity and one of the attempts is to formulate a quantum theory of gravity. \par One of the candidates for quantum gravity is the loop quantum gravity(LQG)\cite{Thiemann2007,Rovelli2004,Giesel2011}. The quantization technique developed in LQG has been applied in the Symmetry reduced Cosmology model, leading to the development of loop quantum cosmology(LQC), which substitutes the classical big bang singularity with a quantum bounce\cite{Bojowald2003,Tsujikawa2004,Bojowald2001,Bojowald2005,Ashtekar2003,Singh2009,Ashtekar2006}. In recent years, loop quantum gravity black hole(LQBH) models have also been constructed to resolve the issue of black hole singularity\cite{Zhang2023,Ashtekar2018,Zhang2022,Zhang2024,Alonso‑Bardaji2022,Gambini2013,Peltola2009}. However, most LQBH models fail to maintain diffeomorphism invariance. Recently, LQBH models maintaining diffeomorphism invariance have been derived by constructing an effective Hamiltonian that preserves covariance\cite{Zhang2025}. The authors continued to extend this approach to derive electrovacuum spacetimes with a cosmological constant\cite{Yang2025}.\par A realistic black hole can never be in an isolated state. Realistic black holes are surrounded by matter and fields. A black hole can also interact with the particles created by vacuum fluctuation, leading to the phenomenon called Hawking radiation. Thus, a realistic black hole always interacts with its surroundings and always lies in a perturbed state. A newly-born black hole formed from gravitational collapse, or from the merger of two compact objects is not in its equilibrium configuration and is therefore perturbed. Once a black hole is perturbed, it tries to reach its equilibrium configuration through some characteristic damped oscillations, the so-called quasinormal modes(QNMs). The spectrum of quasinormal modes of a black hole forms an infinite set of complex angular frequencies. The real part of the frequency is proportional to the oscillation of frequency of the corresponding mode. The negative imaginary part is inversely proportional to the damping time of the mode. If the imaginary part is positive, the oscillation will amplify instead of damping, and the black hole spacetime is not stable under perturbations.\par The spectrum of frequencies of QNMs of black holes does not depend on the initial perturbations, but depends on the specific fundamental properties of a black hole. The spectrum encodes information about the spacetime geometry around a black hole, offering a feasible scheme to measure the black hole parameters\cite{Echeverria1989,Berti2006,Berti2007}. The frequency spectra of QNMs of black holes in different gravity theories are generally different in the sense that different gravity theories usually have different black hole spacetime solutions and different field equations. So the analysis of the QNMs spectrum can be utilized to test different gravity theories\cite{LVK2025,Berti2018b}. \par Quatumn gravity effects modify black holes solutions and hence modify the QNMs spectra of black holes\cite{Lutfuoglu2025,Fu2024,Gong2024,Konoplya2025,Zhu2025,Moreira2023}. In this paper, we will focus on examining the QNMs of scalar perturbations of a charged covariant effective quantum black hole proposed in \cite{Yang2025}. We will not only examine the neutral massless scalar field, but also examine the massive scalar field and the charged scalar field. The black hole solution proposed in \cite{Yang2025} is presented with a cosmological constant. Because we focus on the asymptotically flat black hole spacetime, we will set the cosmological constant off in this work. This paper will be organized as follows. In Section~\ref{sec2}, we will briefly review the charged covariant effective quantum black hole spacetime and have an introduction to the equations of motion in the curved spacetime for different scalar fields.  In Section~\ref{sec3}, we will present an introduction of the numerical method we use in this work to calculate the QNMs. In Section~\ref{sect4}, we will have a detailed investigation on the QNMs of the neutral massless scalar field, the neutral massive scalar field and the charged massless scalar field. Conclusions and discussions will be made in ~\ref{sec5}.

\section{\label{sec2}the charged covariant effective quantum black hole and the equations of motion}
\subsection{\label{sec21}The charged covariant quantum-corrected black hole }
The metric of the charged quantum-corrected black hole that maintains diffeomorphism invariance proposed in \cite{Yang2025} is given by \begin{equation}
    ds^2=-f(r)dt^2+\frac{dr^2}{f(r)} + r^2(d\theta^2+sin^2\theta d\phi^2)\label{1},
\end{equation}
where the metric function$f(r)$ takes the following form\begin{equation}
    f(r)=(1-\frac{2M}{r}+\frac{Q^2}{r^2})\left[1+\frac{\zeta^2}{r^2}(1-\frac{2M}{r}+\frac{Q^2}{r^2})\right]\label{2},
\end{equation}
Here $\zeta$ is the quantum parameter quantifying the deviation of this black hole metric from the classical 
Reissner‑Nordström(RN) black hole metric. 
When $\zeta=0$, the metric reduces to that of the Reissner‑Nordström black hole. $M$ is the ADM mass and in the rest of the paper, we will set $M=1$. $Q$ is the charge of the black hole.\par
The horizons of this black hole locate at the positive real solutions of the equation\begin{equation}
    f(r)=0\label{3},
\end{equation}
which is equivalent to 
\begin{subequations}
    \begin{eqnarray}
     (1-\frac{2}{r}+\frac{Q^2}{r^2})=0,\label{4a}\\
     1+\frac{\zeta^2}{r^2}(1-\frac{2}{r}+\frac{Q^2}{r^2})=0,\label{4b}
    \end{eqnarray}
\end{subequations}
Eq.\eqref{4a} has two real positive solutions $R_{\pm}=1\pm\sqrt{1^{2}-Q^{2}}$ under the condition of $|Q|\leq 1$. $R_{-}$ is exactly the same with the outer horzion of the RN black hole, and $R_{+}$ is exactly the same with the inner horizon of the RN black hole. Eq.\eqref{4b} does not have real positive solutions in the whole $(Q,\zeta)$ parameter space. The conditions under which Eq.\eqref{4b} has real positive solutions are given by 

\begin{widetext}
\begin{equation}
\begin{cases}
\zeta\ge
\sqrt{\dfrac{27 - 36 q^2 + 8 q^4 - \sqrt{\,729 - 1944 q^2 + 1728 q^4 - 512 q^6\,}}{2\left(-1 + q^2\right)}} \\[4pt]
\quad |Q|<1
\end{cases}
\label{eq5}
\end{equation} 
\end{widetext}

\begin{figure}
    \centering
    \includegraphics[width=1\linewidth]{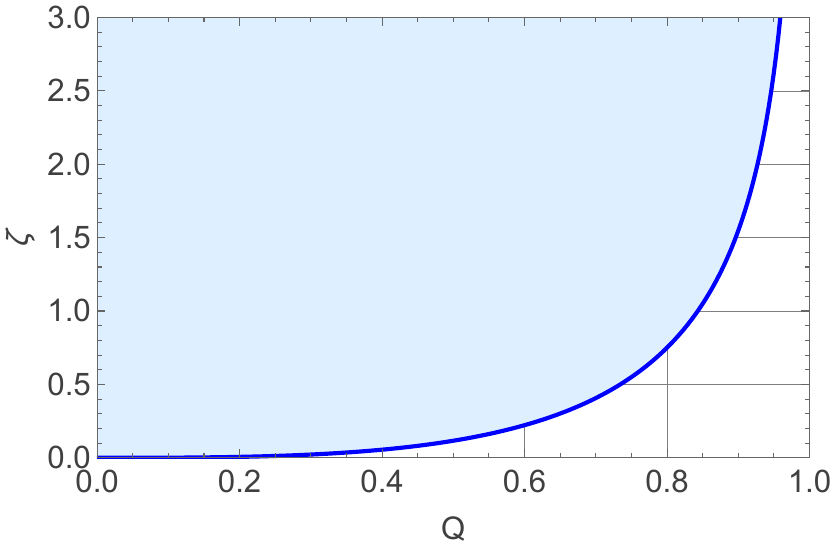}
    \caption{Figure representation of Eq.\eqref{eq5}. Eq.\eqref{4b} has real positive solutions in the filled region.} 
    \label{fig1}
\end{figure}

The parameters in the filled region of Fig.\ref{fig1} satisfy the conditions given by Eq.\eqref{eq5}, and Eq.\eqref{4b} has two real positive solutions in that region, which we will label as $R_{+}^{\prime}$ and  $R_{-}^{\prime}$. The relation between $R_{+}$, $R_{-}$, $R_{+}^{\prime}$ and  $R_{-}^{\prime}$ is 
\begin{equation}
    R_{-}<R_{-}^{\prime}<R_{+}^{\prime}<R_{+}\label{6}
\end{equation}
As we will use Leaver's continued fraction method to calculate the QNMs, this relation is helpful in the construction of the trial solutions of the field equations, which we will see in Section~\ref{sec3}. The range of the quantum parameter $\zeta$ within which we will calculate the QNMs is $0\le\zeta\le3$. This range is chosen according to the constraints on $\zeta$ obtained in \cite{Li2026,Shu2025,Konoplya2025}.

\subsection{\label{22}The equation of motion}
As pointed out in \cite{Konoplya2011}, if there is no back-reaction on the background spacetime, the matter perturbations of a black hole can be studied not only by adding the perturbation terms into the spacetime metric, but also by inducing fields to the spacetime metric. In this case, the QNMs of a matter perturbation of a black hole can be calculated by solving the equation of motion in the background black hole spacetime governing the dynamical evolution of the matter field under specific boundary conditions, although it's nearly impossible to solve the equation analytically. The equation of motion for a scalar field $\Phi$ with mass $m$ in flat spacetime is 
\begin{equation}
    (\eta^{\mu\nu}\partial_\mu\partial_\nu - m^2)\phi = 0,\label{7}
\end{equation}For a massive scalar field minimally-coupled to gravity, we can get its equation of motion simply by replacing the partial derivative in Eq.\eqref{7} with the covariant derivative $\nabla_\mu$ and replacing $\eta^{\mu\nu}$ in Eq.\eqref{7} with $g^{\mu\nu}$. So the equation of motion of a scalar field with mass $m$ takes the form $(g^{\mu\nu}\nabla_\mu\nabla_\nu - m^2)\phi = 0$, which can be rewritten as 
\begin{equation}
\frac{1}{\sqrt{-g}}\partial_\mu\Big(\sqrt{-g}\,g^{\mu\nu}\partial_\nu\phi\Big)-m^2\phi=0,
\label{8}    
\end{equation}
The equation of motion of a charged massless scalar field in curved spacetime is 
\begin{equation}
    D^\mu D_\mu\Phi = 0,\label{9} 
\end{equation}
where $D_\mu = \nabla_\mu - i e A_\mu$ is the \enquote{extended} covariant derivative. $e$ is the charge of the scalar field and $A_\mu$ is the vector potential of the electromagnetic field. Equation \eqref{9} can be rewritten as 
\begin{eqnarray}
    \frac{1}{\sqrt{-g}}\partial_\mu\Big(\sqrt{-g}\,g^{\mu\nu}\big(\partial_\nu\Phi - ie A_\nu \Phi\big)\Big)\nonumber\\
-ie A^\mu \partial_\mu\Psi
-e^2 A^\mu A_\mu\Phi = 0,\label{10}
\end{eqnarray}

The vector potential of the electromagnetic field around a RN like black hole is 
\begin{equation}
    A_\mu=\left(-\frac{Q}{r},\;0,\;0,\;0\right), \label{11}
\end{equation}
Put Eq.\eqref{11} into Eq.\eqref{10}, we can get the equation of motion of the charged massless scalar field around a charged covariant quantum-corrected black hole as follow 

\begin{eqnarray}
\frac{1}{\sqrt{-g}} \partial_\mu \left( \sqrt{-g} g^{\mu\nu} \partial_\nu \phi \right)
- \frac{i e Q}{\sqrt{-g}} \partial_t \left( \sqrt{-g} \frac{g^{tt}}{r} \phi \right)\nonumber\\
- \frac{i e Q g^{tt}}{r} \partial_t \phi
- \frac{e^2 Q^2 g^{tt}}{r^2} \phi = 0,\label{12}
\end{eqnarray}
In order to separate variables and to reduce equations of motion to ordinary differential equations, we can choose $\Phi$ as 
\begin{equation}
    \Phi(t, r, \theta, \varphi) = e^{-i\omega t} Y_{\ell}(\theta, \varphi) \frac{\phi(r)}{r},\label{13}
\end{equation}
Putting Eq.\eqref{13} into equations of motion and using the \enquote{tortoise} coordinate $r_*$, which is defined as $dr_*\equiv\displaystyle\frac{dr}{f(r)}$, Both Eq.\eqref{8} and Eq.\eqref{12} can be reduced to a radial equation in the following form\begin{equation}
\frac{d^2\phi}{dr_*^2} + \left( \omega^2 - V(r) \right) \phi = 0,
\label{14}    
\end{equation}
where $V(r)$ is the effective potential.\par
For the neutral massive scalar field, its effective potential takes the following form
\begin{equation}
V(r) = \frac{l(l+1) f(r)}{r^2} + \frac{f(r) f'(r)}{r}+m^2f(r),
\label{15}
\end{equation}
and for the charged massless scalar field, its effective potential takes the following from 
\begin{equation}
V(r) =\frac{\ell(\ell+1) f(r)}{r^2} + \frac{f(r) f'(r)}{r} +\frac{2 e Q \omega}{r} -\frac{e^2 Q^2}{r^2},
\label{16}
\end{equation}
To find the frequency spectra of the QNMs, we need to solve Eq.\eqref{14} using the following boundary conditions
\begin{equation}
\phi(r_*) \sim e^{\pm i \omega r_*}, \qquad r_* \to \pm \infty,
\label{17}\end{equation}
Thus, we require that at the event horizon of the black hole, the wave is purely ingoing and at the spatial infinity, the wave is purely outgoing.
\section{\label{sec3}The continued fraction method}
To solve an equation like Eq.\eqref{14} is virtually impossible, and various numerical approaches have been developed to calculate QNMs, including the WKB method\cite{Ferrari1984,Schutz1985,Iyer1987,Iyer1987b,2003PhRvD..68b4018K},the asymptotic iteration method (AIM)\cite{Ciftci2005,Cho2010,Cho2012}, the Horowitz-Hubeny method\cite{Horowitz2000}, the Chebyshev pseudospectral
method(PSM)\cite{Boyd2001,Jansen2017} and the continued fraction method\cite{Leaver1985}. Among all those numerical methods, the continued fraction method is the most reliable one and gives us the most accurate result\cite{Konoplya2025}. Because of this, we use the continued fraction method in this study. For faster convergence, we adopt the improved scheme proposed by Nollert\cite{Nollert1993} in its general form developed in \cite{Zhidenko2006}.\par We now show explicitly how to apply the continued fraction method. 
If we put the usual radial coordinate $r$ back, Eq.\eqref{14} can be rearranged into the following form
\begin{equation}
\left(\frac{d^2}{dr^2}+\frac{d}{dr}\frac{f'(r)}{f(r)}+\frac{\left(\omega^2-V(r)\right)}{f^2(r)}\right)\phi=0,
\label{18}\end{equation}
If the parameters $Q$ and $\zeta$ satisfy Eq.\eqref{eq5}, Eq.\eqref{18} has six singular points $\{0, R_{-},R_{-}^{\prime},R_{+}^{\prime},R_{+},\infty\}$, and if not, Eq.\eqref{18} has four singular points $\{0,R_{-},R_{+},\infty\}$. If we introduce a new variable 
\begin{equation}
    z=\dfrac{r - R_{+}}{r - r_-},\label{19}
\end{equation} 
where $r_-$ should be a real positive number smaller than $R_{+}$. The domain outside the black hole $(R_{+},\infty)$ can be mapped into the domain $(0,1)$. We can write the solution of Eq.\eqref{18} in a series form 
\begin{equation}
\psi(r)=e^{i\Omega r}(r-r_-)^\sigma\left(\frac{r-R_+}{r-r_-}\right)^{-ia}\sum_{k=0}^{\infty} b_k \left(\frac{r-R_+}{r-r_-}\right)^k,\label{20}
    \end{equation}
where $\Omega$, $\sigma$ and $a$ are chosen such that the boundary condition \eqref{17} is satisfied. 
If after the mapping with the use of Eq.\eqref{19}, all the singular points of Eq.\eqref{18} are outside the domain $(0,1)$, the series Eq.\eqref{20} is convergent at $z=1$ if and only if $\omega$ is the eigenfrequency of Eq.\eqref{18}. This can be done by choosing appropriate $r_-$ in Eq.\eqref{19}. If the parameters $Q$ and $\zeta$ satisfy Eq.\eqref{eq5}, $r_-$ is chosen to be $R_{+}^{\prime}$. If not, $r_-$ is chosen to be $R_{-}$.\par
Substituting Eq.\eqref{20} into Eq.\eqref{18}, we can obtain an $N$-term recurrence relation for the expansion coefficient $b_i$ 
\begin{equation}
\sum_{m=0}^{\min(N-1,n)} \alpha_{m,n}^{(N)} b_{n-m}=0,\quad \text{for } n>0,\label{21}
\end{equation}
where the coefficients $\alpha_{m,n}^{N}$ depend on the QNMs frequency $\omega$.
Because the number of terms $N$ we obtain in this study is larger than $3$, for example, for the massive scalar field, we have a 15-term recurrence relation, we need to use \textit{Gaussian eliminations} to reduce the number of terms to 3.
\par 
The Gaussian eliminations procedure can be derived from the recurrence relation Eq.\eqref{21}. Suppose we have a $k+1$-term recurrence relation
\begin{equation}
\sum_{m=0}^{\min(k,n)} \alpha_{m,n}^{(k+1)} b_{n-m}=0,\label{22}
\end{equation}
The corresponding $k$-term recurrence relation is 
\begin{equation}
\sum_{m=0}^{\min(k-1,n)} \alpha_{m,n}^{(k)} b_{n-m}=0,\label{23}
\end{equation}
If $n<k$, comparing Eq.\eqref{22} with Eq.\eqref{23}, we have 
\begin{equation}
    \alpha_{m,n}^{(k+1)} = \alpha_{m,n}^{(k)},\label{24}
\end{equation}
If $n\ge k$, Eq.\eqref{23} can be rewritten as 
\begin{equation}
\sum_{m=0}^{k-1} \alpha_{m,n}^{(k)} b_{n-m}=0,\label{25}
\end{equation}
Replacing $n$ with $n-1$ and replacing $m$ with $m-1$ in Eq.\eqref{25}, we have 
\begin{equation}
\sum_{m=1}^{k} \alpha_{m-1,n-1}^{(k)} b_{n-m}=0,\label{26}
\end{equation}
Multiplying Eq.\eqref{26} by $\alpha_{k,n}^{(k+1)} / \alpha_{k-1,n-1}^{(k)}$, we have 
\begin{equation}
\dfrac{\alpha_{k,n}^{(k+1)}}{\alpha_{k-1,n-1}^{(k)}} \sum_{m=1}^{k} \alpha_{m-1,n-1}^{(k)} b_{n-m} = 0,\label{27}
\end{equation}
Subtracting Eq.\eqref{27} from Eq.\eqref{22}, we have 
\begin{equation}
    \begin{aligned}
\alpha_{0,n}^{(k+1)} b_n
&+ \sum_{m=1}^{k-1}
\left(
\alpha_{m,n}^{(k+1)}
- \frac{\alpha_{k,n}^{(k+1)}}{\alpha_{k-1,n-1}^{(k)}}
\alpha_{m-1,n-1}^{(k)}
\right) b_{n-m} \\
&+ \left(
\alpha_{k,n}^{(k+1)}
- \frac{\alpha_{k,n}^{(k+1)}}{\alpha_{k-1,n-1}^{(k)}}
\alpha_{k-1,n-1}^{(k)}
\right) b_{n-k}
= 0,
\end{aligned}\label{28}
\end{equation}
Comparing Eq.\eqref{28} with Eq.\eqref{25}, we have 
\begin{equation}
    \begin{cases}
\alpha_{m,n}^{k+1} = \alpha_{m,n}^{k}, & \text{for } m=0,\\[6pt]
\alpha_{m,n}^{(k)} = \alpha_{m,n}^{(k+1)} - \dfrac{\alpha_{k,n}^{(k+1)} \alpha_{m-1,n-1}^{(k)}}{\alpha_{k-1,n-1}^{(k)}}, & \text{for } m\neq 0.
\end{cases}\label{29}
\end{equation}
Combining Eq.\eqref{29} and Eq.\eqref{24}, the Gaussian eliminations procedure is given by 
\begin{equation}
    \begin{cases}
\alpha_{m,n}^{(k+1)} = \alpha_{m,n}^{(k)}, & \text{for } m=0 \text{ or } n<m,\\[8pt]
\alpha_{m,n}^{(k)} = \alpha_{m,n}^{(k+1)} - \dfrac{\alpha_{k,n}^{(k+1)} \alpha_{m-1,n-1}^{(k)}}{\alpha_{k-1,n-1}^{(k)}}, & \text{for } m\neq 0 \text{ and } n\ge m.
\end{cases}\label{30}
\end{equation}
\par If we have a $N$-term recurrence relation, by iterating Eq.\eqref{30} $N-3$ times, we finally have a $3$-term recurrence relation
\begin{subequations}\label{31}
\begin{align}
\alpha_{0,n}^{(3)} b_n + \alpha_{1,n}^{(3)} b_{n-1} + \alpha_{2,n}^{(3)} b_{n-2} &= 0,\quad n>1 \label{31a}\\
\alpha_{0,1}^{(3)} b_1 + \alpha_{1,1}^{(3)} b_0 &= 0 \label{31b}
\end{align}
\end{subequations}    
From Eq.\eqref{31b}, we have 
\begin{equation}
    \dfrac{b_1}{b_0} = -\dfrac{\alpha_{1,1}^{(3)}}{\alpha_{0,1}^{(3)}}\label{32}
\end{equation}
Substituting Eq.\eqref{32} into Eq.\eqref{31a}, we have 
\begin{equation}
    -\frac{\alpha_{1,1}^{(3)}}{\alpha_{0,1}^{(3)}}
=
-\cfrac{\alpha_{2,2}^{(3)}}
{\alpha_{1,2}^{(3)}
- \cfrac{\alpha_{0,2}^{(3)}\,\alpha_{2,3}^{(3)}}
{\alpha_{1,3}^{(3)}
- \cfrac{\alpha_{0,3}^{(3)}\,\alpha_{2,4}^{(3)}}
{\alpha_{1,4}^{(3)}
- \cdots
}
}
},\label{33}
\end{equation}
which can be rearranged as 
\begin{equation}
    0=\alpha_{1,1}^{(3)}
- \cfrac{\alpha_{0,1}^{(3)}\,\alpha_{2,2}^{(3)}}
{\alpha_{1,2}^{(3)}
- \cfrac{\alpha_{0,2}^{(3)}\,\alpha_{2,3}^{(3)}}
{\alpha_{1,3}^{(3)}
- \cdots
}
},\label{34}
\end{equation}
If we inverted Eq.\eqref{34} $n$ times, we have 
\begin{align}
&\alpha_{1,n+1}^{(3)}
- \cfrac{\alpha_{2,n}^{(3)}\,\alpha_{0,n-1}^{(3)}}
{\alpha_{1,n-1}^{(3)}
- \cfrac{\alpha_{2,n-1}^{(3)}\,\alpha_{0,n-2}^{(3)}}
{\alpha_{1,n-2}^{(3)}
- \cdots
- \cfrac{\alpha_{2,2}^{(3)}\,\alpha_{0,1}^{(3)}}
{\alpha_{1,1}^{(3)}}
}
}
\nonumber\\
&\hphantom{\alpha_{1,n+1}^{(3)}} 
= \cfrac{\alpha_{0,n+1}^{(3)}\,\alpha_{2,n+2}^{(3)}}
{\alpha_{1,n+2}^{(3)}
- \cfrac{\alpha_{0,n+2}^{(3)}\,\alpha_{2,n+3}^{(3)}}
{\alpha_{1,n+3}^{(3)}
- \cdots
}
},\label{35}
\end{align}

Eq.\eqref{35} can be solved numerically. Because of the infinite continued fraction on the right-hand side, we have to truncate its length to a finite number $N$. The choice of $N$ must satisfy the requirement that an increase in $N$ does not change the numerical solution within desired precision. The most stable root of Eq.\eqref{35} is the $n$-th overtone QNM. So generally, we have to use the $n$ times inverted equation to find the $n$-th overtone. 
\par In order to have a faster convergence, we adopt the improved scheme proposed by Nollert \cite{Nollert1993} in its general form developed in \cite{Zhidenko2006}. Without the improvement, if we truncate the length of the infinite continued fraction to $N$, we usually set $b_{N+1}/b_N=0$. The idea of the improvement is that for large $n$, we can expand 
\begin{equation}
    R(n)\equiv \dfrac{b_n}{b_{n-1}}=- \cfrac{\alpha_{2,n+1}^{(3)}}
{\alpha_{1,n+1}^{(3)}
- \cfrac{\alpha_{0,n+1}^{(3)}\,\alpha_{2,n+2}^{(3)}}
{\alpha_{1,n+2}^{(3)}
- \cdots
}
}\label{36}
\end{equation}
as \begin{equation}
R_n(\omega) \equiv C_0(\omega) + \dfrac{C_1(\omega)}{\sqrt{n}} + \dfrac{C_2(\omega)}{n} + \cdots,\label{37}
\end{equation} If we divide Eq.\eqref{21} by $b_{n-N+1}$, and use the definition \eqref{36}, we can have the following equation
\begin{equation}
\sum_{m=0}^{N-2} \alpha_{m,n}^{(N)} \prod_{j=m}^{N-2} R_{n-j} + \alpha_{N-1,n}^{(N)}=0,
\end{equation}\label{38}
For large $n$, the coefficients $\alpha_{m,j}^{(N)}\propto n^2$, we have 
\begin{equation}
\lim_{n\to\infty} 
\dfrac{1}{n^2}\alpha_{m,n}^{(N)} C_0^{N-1-m} = 0,\label{39}
\end{equation}
In general case, Eq.\eqref{39} has $N-1$ roots. If the series Eq.\eqref{20} has a unit radius of convergence, one of the roots is always $C_0=1$, and $C_0$ is chosen to be 1. After fixing $C_0=1$, one can find equations for other coefficients in Eq.\eqref{37} and other coefficients are thus determined. The expansion \eqref{37} are a good approximation for a sufficient large $n$. 
\section{\label{sect4} the QNMs under the influence of $Q$ and $\zeta$}
\subsection{\label{sect41} Neutral massless scalar field }
\begin{figure}[htbp]
  \centering
  \begin{minipage}{0.48\textwidth}
    \centering
    \includegraphics[width=\linewidth]{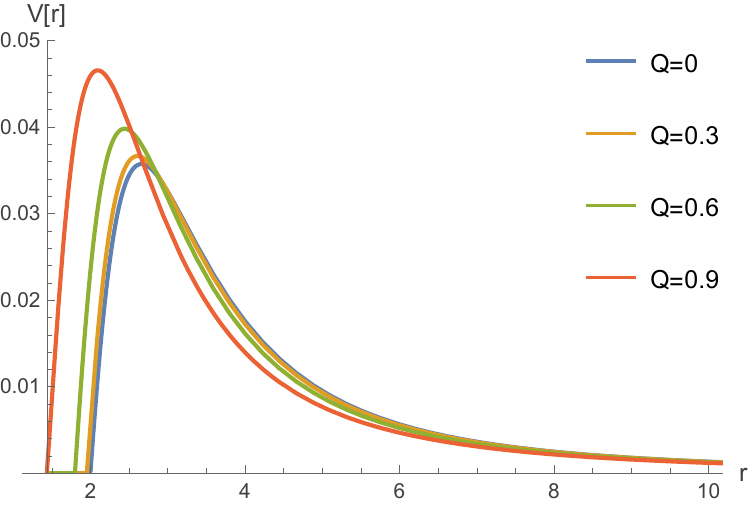}
  \end{minipage}
  \hfill
  \begin{minipage}{0.48\textwidth}
    \centering
    \includegraphics[width=\linewidth]{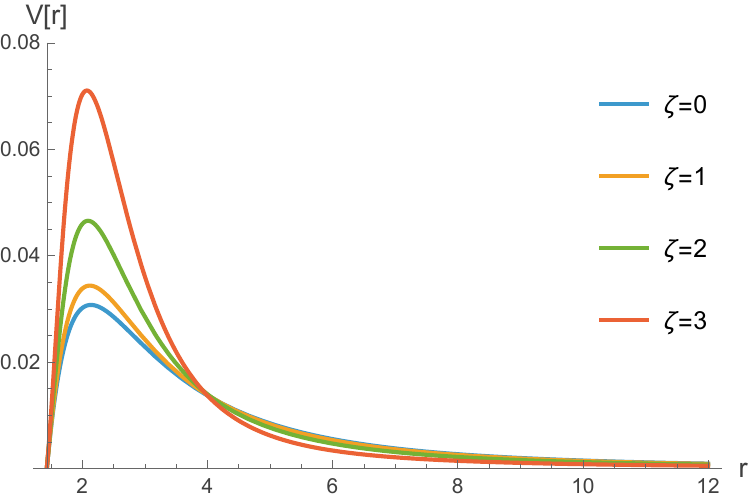}
  \end{minipage}
  \caption{The effective potential of neutral massless scalar field as a function of $r$ with $l=0$. In the upper panel, $\zeta$ is fixed at $\zeta=2$, while  charge $Q$ is varied. In the lower panel, charge $Q$ is fixed at $Q=0.9$, while $\zeta$ is varied}\label{fig2}
\end{figure}

\begin{figure}[htbp]
  \centering
  \begin{minipage}{0.48\textwidth}
    \centering
    \includegraphics[width=\linewidth]{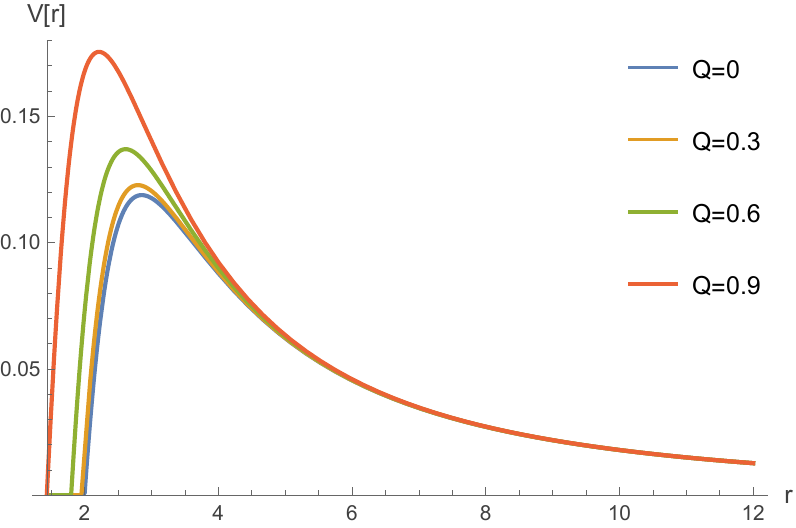}
  \end{minipage}
  \hfill
  \begin{minipage}{0.48\textwidth}
    \centering
    \includegraphics[width=\linewidth]{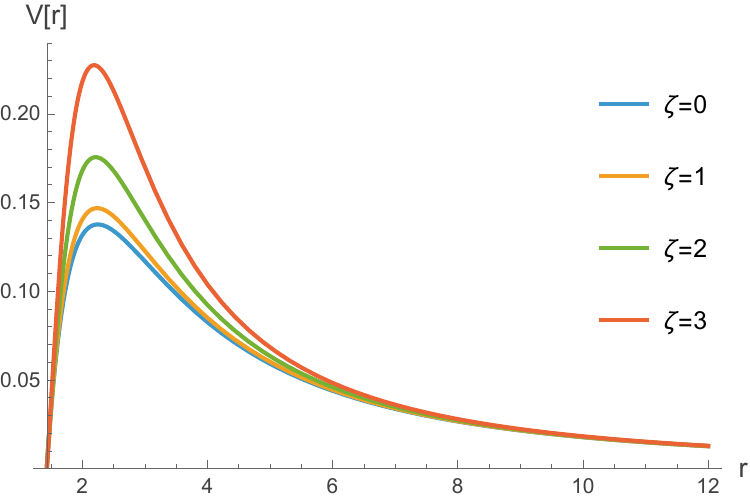}
  \end{minipage}
  \caption{The effective potential of neutral massless scalar field as a function of $r$ with $l=1$. In the upper panel, $\zeta$ is fixed at $\zeta=2$, while  charge $Q$ is varied. In the lower panel, charge $Q$ is fixed at $Q=0.9$, while $\zeta$ is varied}\label{fig3}
\end{figure}

In this study, we examine the QNMs of neutral massless scalar field, neutral massive scalar field and charged massless scalar field. In this subsection, we will have a detailed investigation on the QNMS of the neutral massless field. The effective potentials for neutral massless scalar filed for multipole number $l=0$ and $l=1$ are shown in Fig.\ref{fig2} and Fig.\ref{fig3} respectively. The potential remains positive in the whole parameter space, indicating that the charged covariant quantum-corrected black hole under consideration is stable under neutral massless scalar perturbation. With the quantum parameter turned on, the maximum value of the effective potential increases as $Q$ increases, which is consistent with the RN black hole case. For fixed $Q$, the maximum value of the effective potential also increases as the quantum parameter increases. We also note that $dV_{max}/d\zeta$ increases dramatically as $\zeta$ increases, where $V_{max}$ denotes the maximum value of the effective potential. So does $dV_{max}/dQ$ when $Q$ increases. This may indicate that a change in a large parameter $\zeta$ or $Q$ may result in a more significant change in the QNMs spectrum than the same change in a small parameter. 
\begin{figure*}[htb]
\centering
\subfloat[]{\includegraphics[width=0.42\linewidth]{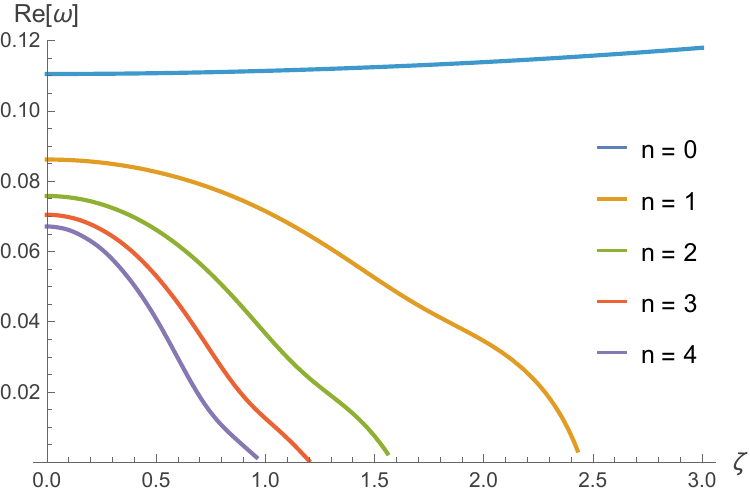}}
\hspace{0.02\linewidth}
\subfloat[]{\includegraphics[width=0.42\linewidth]{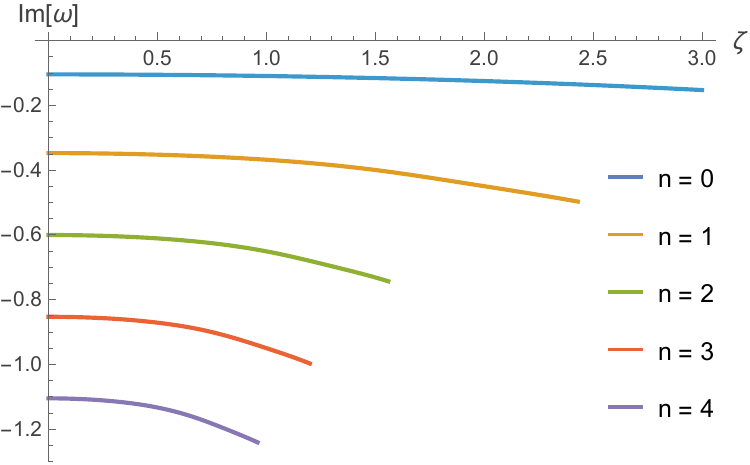}}
\vspace{4pt}
\subfloat[]{\includegraphics[width=0.42\linewidth]{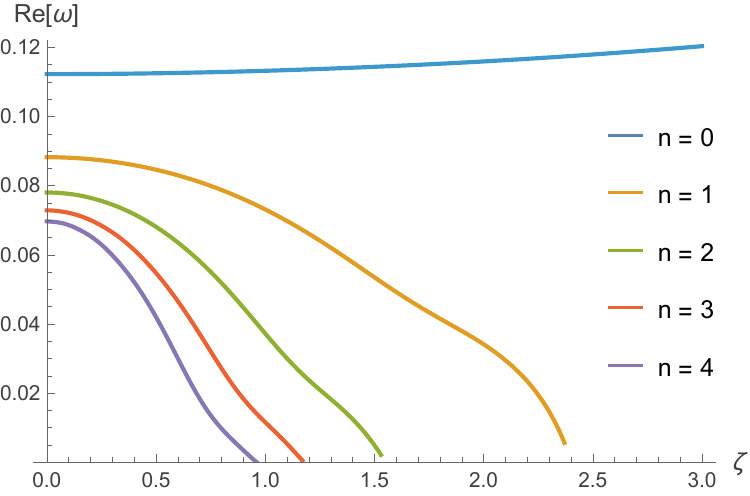}}
\hspace{0.02\linewidth}
\subfloat[]{\includegraphics[width=0.42\linewidth]{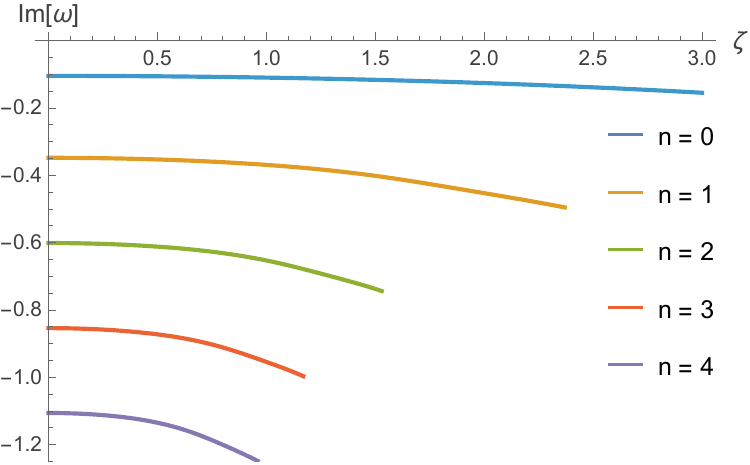}}
\vspace{4pt}
\subfloat[]{\includegraphics[width=0.42\linewidth]{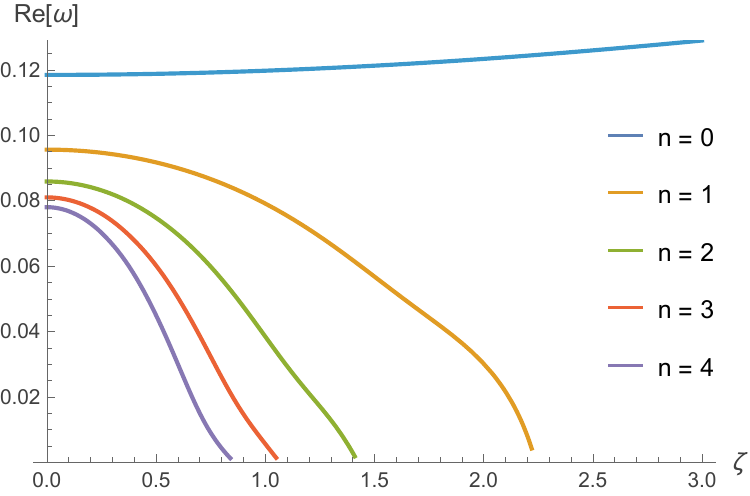}}
\hspace{0.02\linewidth}
\subfloat[]{\includegraphics[width=0.42\linewidth]{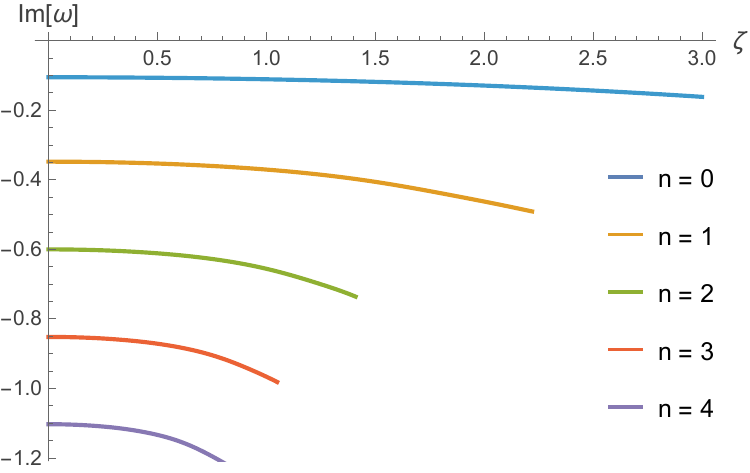}}
\caption{The QNMs of the fundamental mode and the first four overtones as a function of $\zeta$ for $l=0$ when the charge $Q$ is small. The charges in the top, middle, and bottom panels are $0$, $0.3$, $0.6$, respectively}
\label{fig4}
\end{figure*}
\begin{table*}[htbp]
\centering
\caption{Quasinormal frequencies of the fundamental mode of charged and uncharged massless scalar field: $l=0$, $Q=0.9999$.}
\label{tab1}
\begin{tabular*}{\textwidth}{@{\extracolsep{\fill}} cccc @{}}
\toprule
$\zeta$ & $e=0$ & $e=0.3$ & $e=0.6$ \\
\midrule
$0$ & $0.133459 - 0.0958438\,i$ & $0.300363 - 0.0124434\,i$ & $0.595933 - 0.00721299\,i$ \\
$1$ & $0.136828 - 0.103587\,i$ & $0.300289 - 0.0125072\,i$ & $0.596141 - 0.00712849\,i$ \\
$2$ & $0.146361 - 0.128599\,i$ & $0.300107 - 0.0126366\,i$ & $0.596828 - 0.00716096\,i$ \\
$3$ & $0.162608 - 0.175718\,i$ & $0.299891 - 0.0127604\,i$ & $0.597573 - 0.00779621\,i$ \\
\bottomrule
\end{tabular*}
\end{table*}

\begin{table}[htbp]
\centering
\caption{Quasinormal frequencies of the fundamental mode with $l=3$ for different field charge $e$ at $Q=0.9999$.}
\label{table2}
\begin{tabular*}{\columnwidth}{@{\extracolsep{\fill}} c c c @{}}
\toprule
$e$ & $\zeta=0$ & $\zeta=1$ \\
\midrule
$0.0$ & $0.876037 - 0.0885890\,i$ & $0.902584 - 0.0967317\,i$ \\
$0.1$ & $0.926722 - 0.0884854\,i$ & $0.953266 - 0.0966285\,i$ \\
$0.3$ & $1.03229 - 0.0876408\,i$  & $1.05848 - 0.0957108\,i$  \\
$0.6$ & $1.20114 - 0.0847992\,i$  & $1.22594 - 0.0925757\,i$  \\
\bottomrule
\end{tabular*}
\end{table}
\begin{figure*}[htb]
\centering
\subfloat[]{\includegraphics[width=0.29\linewidth]{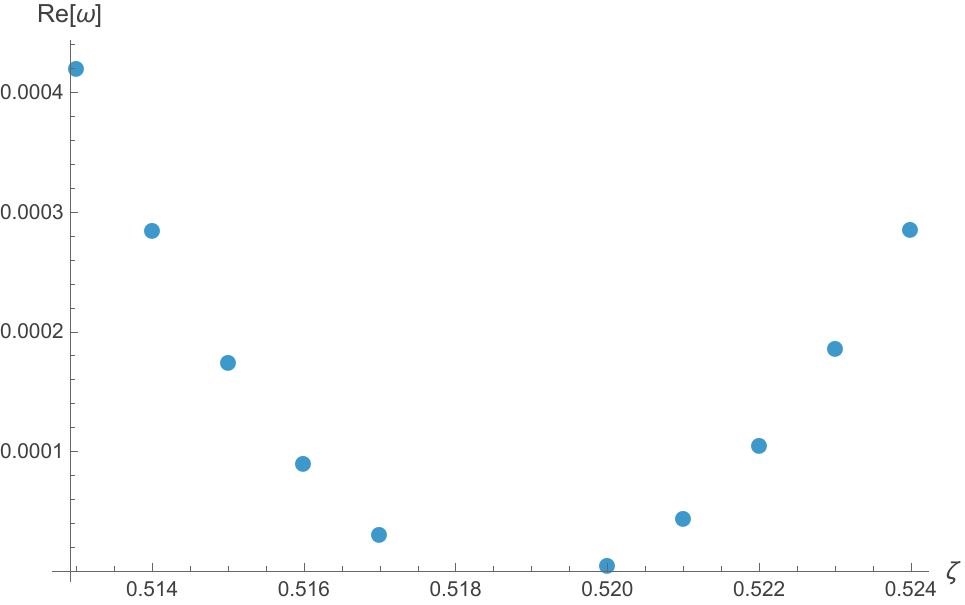}}
\hspace{0.02\linewidth}
\subfloat[]{\includegraphics[width=0.29\linewidth]{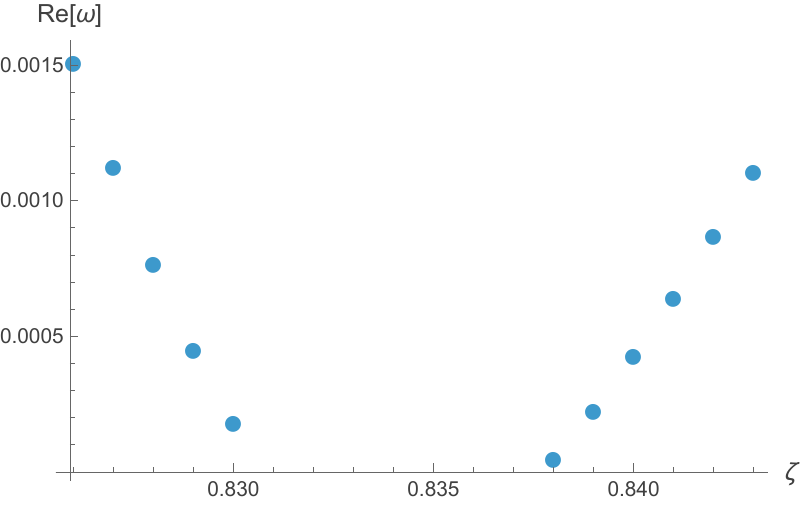}}
\hspace{0.02\linewidth}
\subfloat[]{\includegraphics[width=0.29\linewidth]{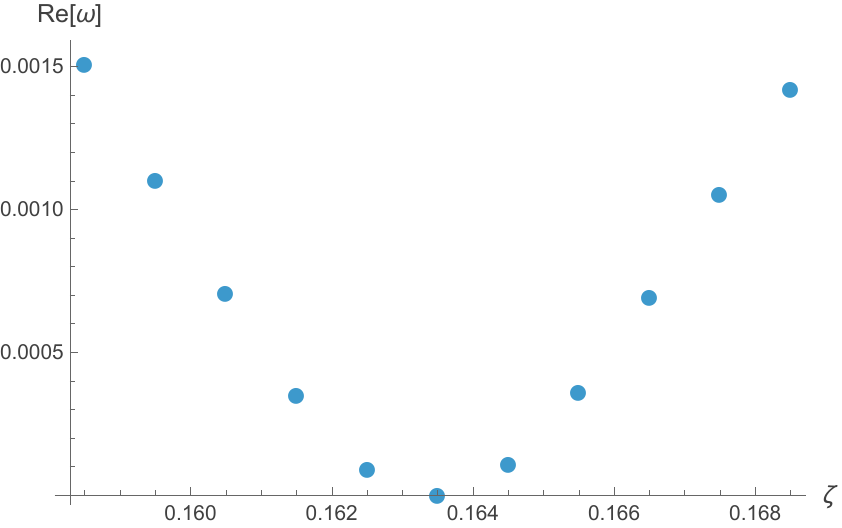}}

\caption{Enlargement of the region where the real part of the quasinormal frequency approaches zero. The left panel corresponds to the third overtone for charge $0.88$, the middle panel to the second overtone for charge $0.9$, and the right panel to the third overtone for charge $0.9$.}
\label{fig6}
\end{figure*}

\begin{figure*}[htb]
\centering
\subfloat[]{\includegraphics[width=0.42\linewidth]{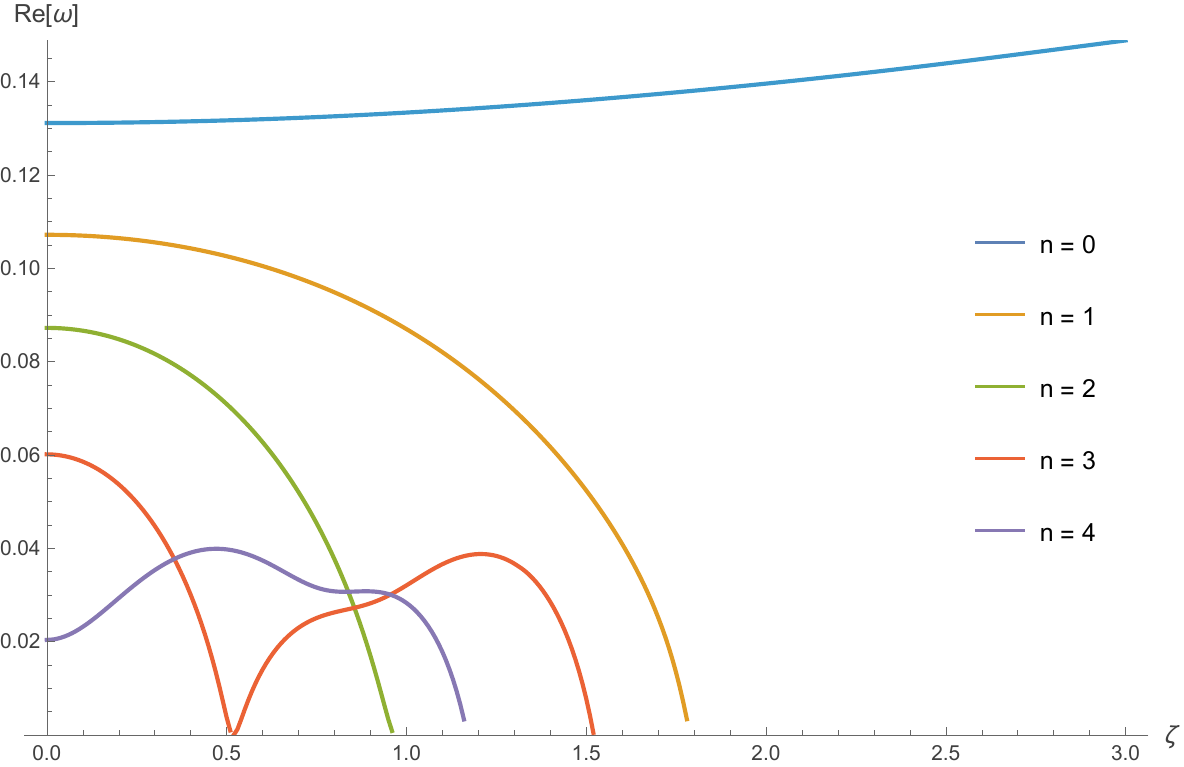}}
\hspace{0.02\linewidth}
\subfloat[]{\includegraphics[width=0.42\linewidth]{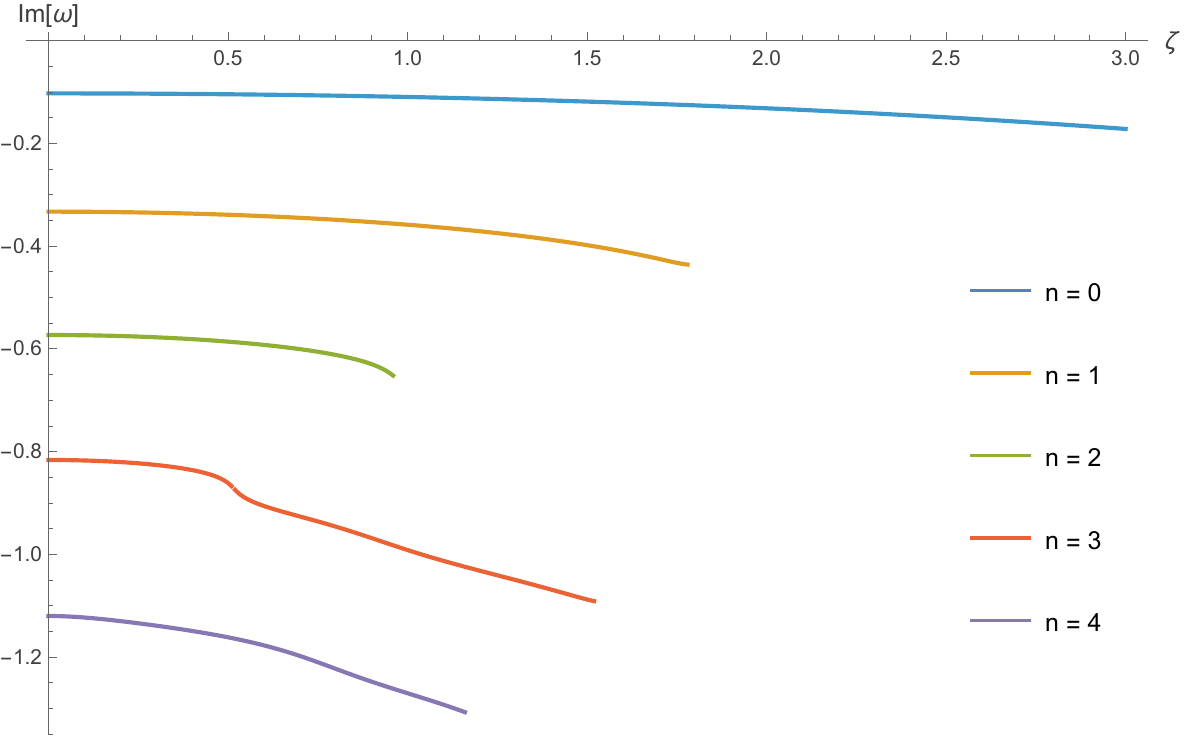}}
\vspace{4pt}
\subfloat[]{\includegraphics[width=0.42\linewidth]{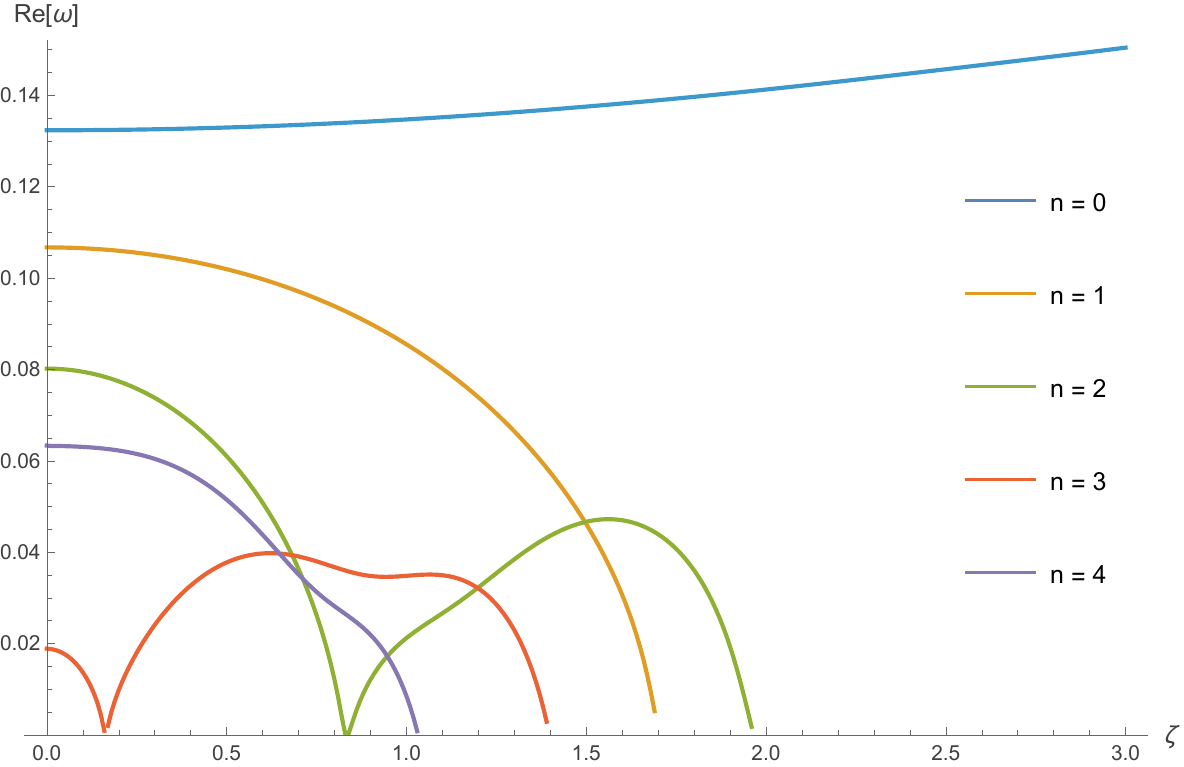}}
\hspace{0.02\linewidth}
\subfloat[]{\includegraphics[width=0.42\linewidth]{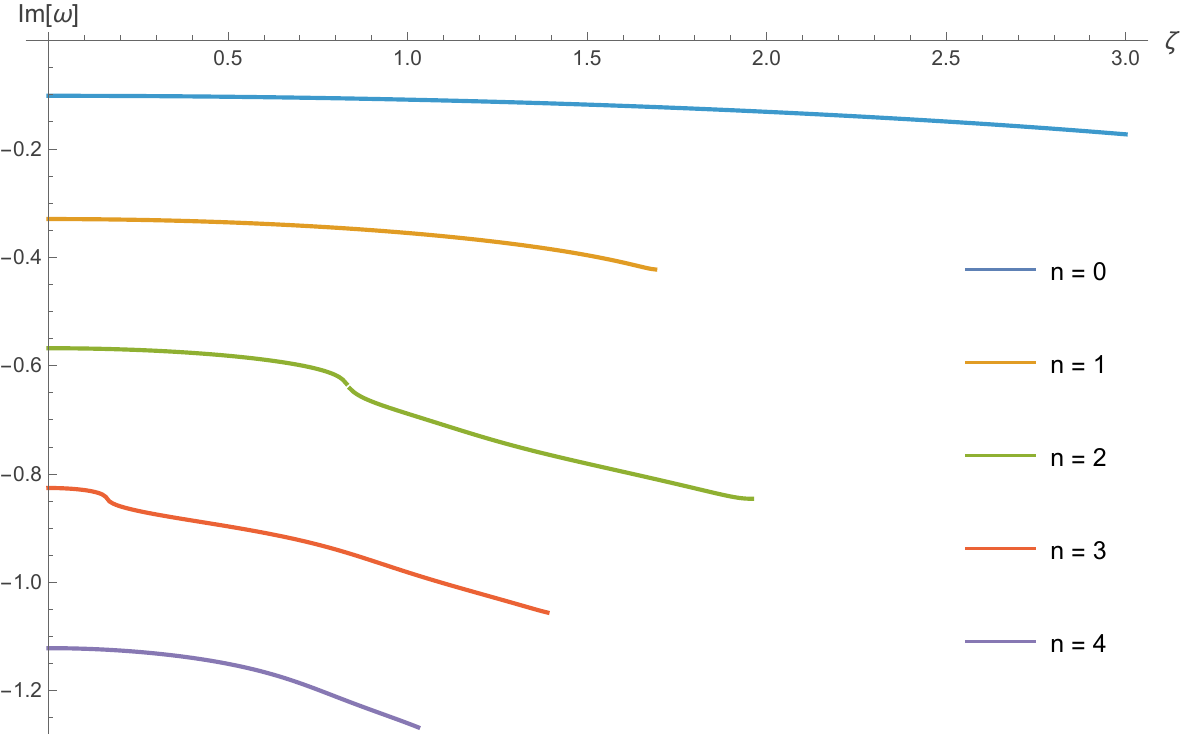}}
\vspace{4pt}
\subfloat[]{\includegraphics[width=0.42\linewidth]{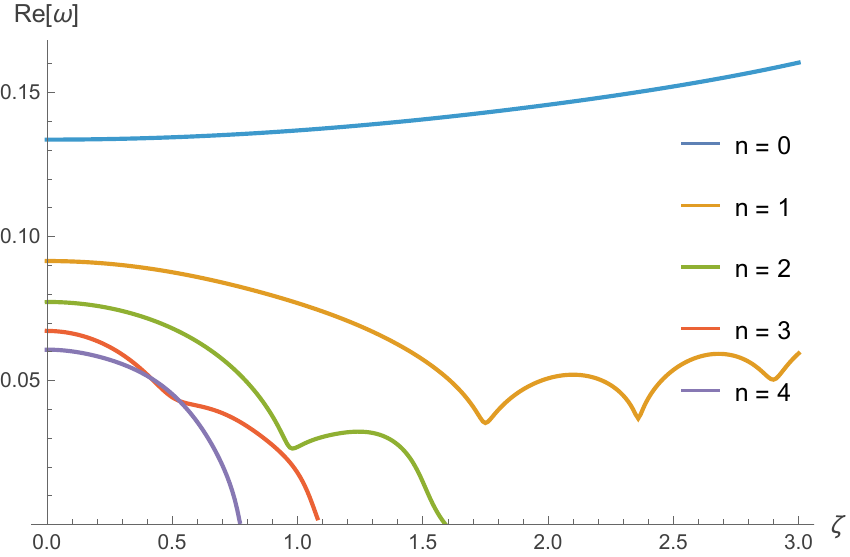}}
\hspace{0.02\linewidth}
\subfloat[]{\includegraphics[width=0.42\linewidth]{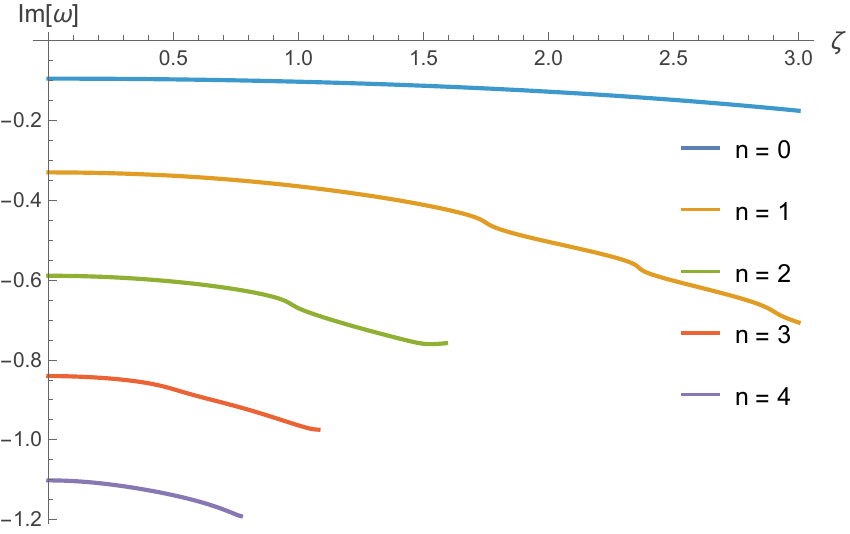}}
\vspace{4pt}
\subfloat[]{\includegraphics[width=0.42\linewidth]{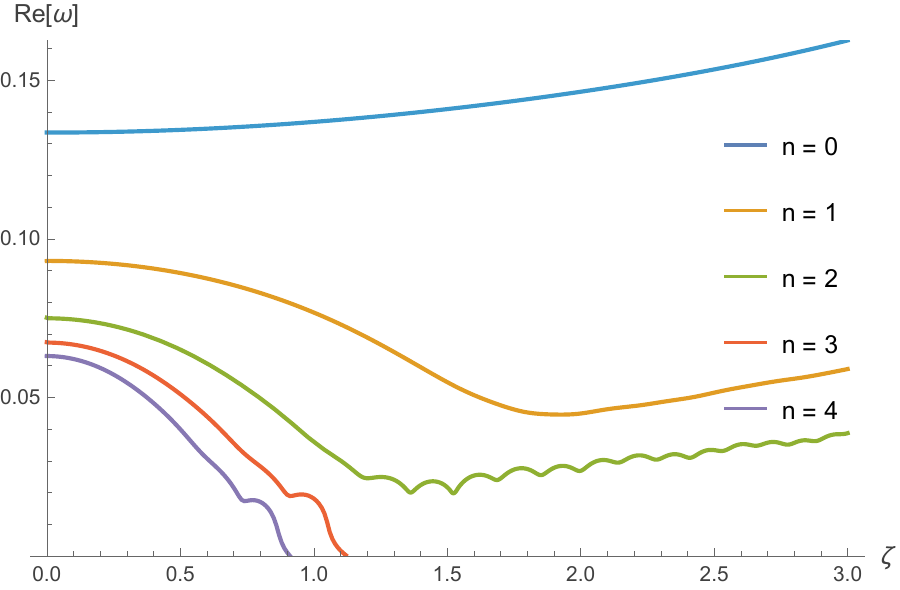}}
\hspace{0.02\linewidth}
\subfloat[]{\includegraphics[width=0.42\linewidth]{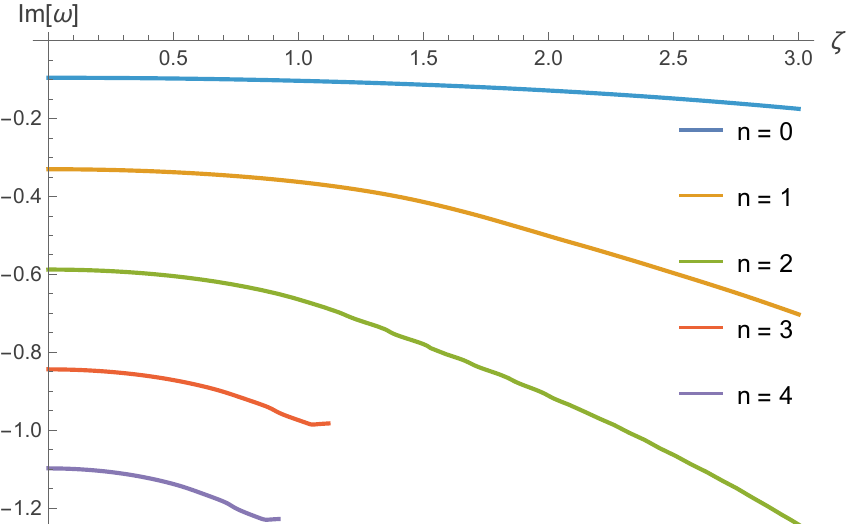}}
\caption{The QNMs of the fundamental mode and the first four overtones as a function of $\zeta$ for $l=0$ when the charge $Q$ is large. The charges in the first, second, third, and fourth rows are $0.88$, $0.9$, $0.99$, and $0.999$, respectively.}
\label{fig5}
\end{figure*}

\par Fig.\ref{fig4} and Fig.\ref{fig5} show how the fundamental mode and the four overtones change with the quantum parameter$\zeta$ when the charge $Q$ is fixed. For small charges, their behaviors are qualitatively the same as that in the uncharged black hole case: for fundamental modes, the real part of the quasinormal frequency increases monotonically as $\zeta$ increases while for the overtones, the real part of the quasinormal frequency decreases monotonically with the increase of $\zeta$ and the real part of the quasinormal frequency approaches zero at some critical $\zeta$, and then disappears from the spectrum; for the fundamental mode and the overtones, the imaginary part decreases monotonically.
The vanishing of oscillation frequency of QNMS when the quantum correction is made is also observed in \cite{Zinhailo2024b}. The change in the fundamental mode is mild over the whole range of $\zeta$ while the change in the overtones is sharp. This phenomenon is called the “outburst of overtones” and is explained by the fact that the overtones are highly sensitive to even slight change in the near-horizon zone while the fundamental mode is primarily determined by the behavior of the effective potential near the peak of the potential barrier\cite{Konoplya2024}. For large charges, the behavior of the overtones under the influence of the quantum parameter $\zeta$ is qualitatively different from that in the uncharged black hole case while the fundamental mode remains qualitatively the same. As we can see in Fig.\ref{fig5}, as the charge increases, non-monotonic behaviors of the real part of QNMS under the influence of $\zeta$ emerge and very strangely, some of the modes can even bounce back after they reach zero and then re-approach zero and finally disappear from the spectrum.

Fig.\ref{fig6} shows the enlargement view of the region where the real part of the frequency goes zero. When the charge is $0.88$, the third overtone approaches zero at some critical value of $\zeta$ and the mode disappears for a while before it reemerges at a larger value of $\zeta$. We see a similar behavior in the second overtone when the charge is $0.9$. It seems that for the third overtone with $Q=0.9$, the mode immediately bounces back after its real part reaches zero and it never disappears from the spectrum. As the extremal charge is approached, the real parts of some of the modes cease to approach zero in the whole range of $\zeta$ under consideration. When the charge is $0.99$, the real part of the second overtone ceases to tend to zero, while when the charge is $0.999$, the real parts of both the second and the third overtone cease to tend to zero. 

The observation that for small charges, the behaviors of overtones as the quantum parameter changes are qualitatively the same as that in the uncharged black hole case, while for large charges, the behaviors of overtones are qualitatively different from that in the uncharged case can also be seen in the case of $l=1$, as we can see in Fig.\ref{7}.
\begin{figure*}[htb]
\centering
\subfloat[]{\includegraphics[width=0.42\linewidth]{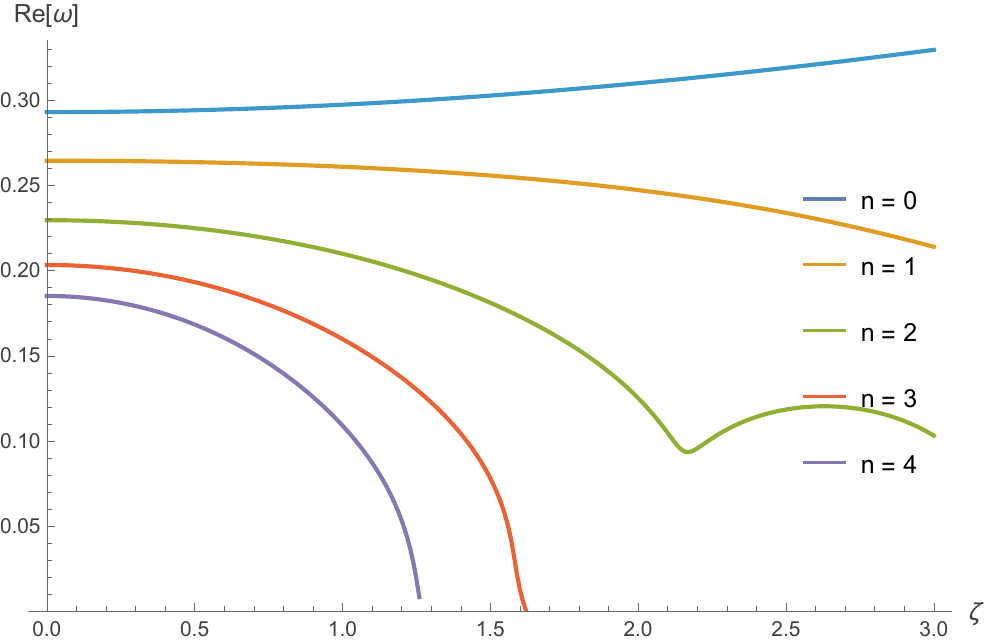}}
\hspace{0.02\linewidth}
\subfloat[]{\includegraphics[width=0.42\linewidth]{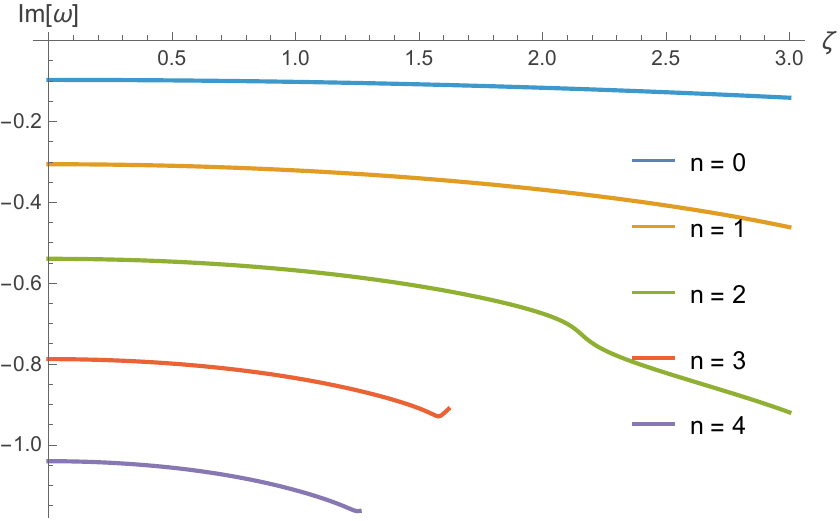}}
\vspace{4pt}
\subfloat[]{\includegraphics[width=0.42\linewidth]{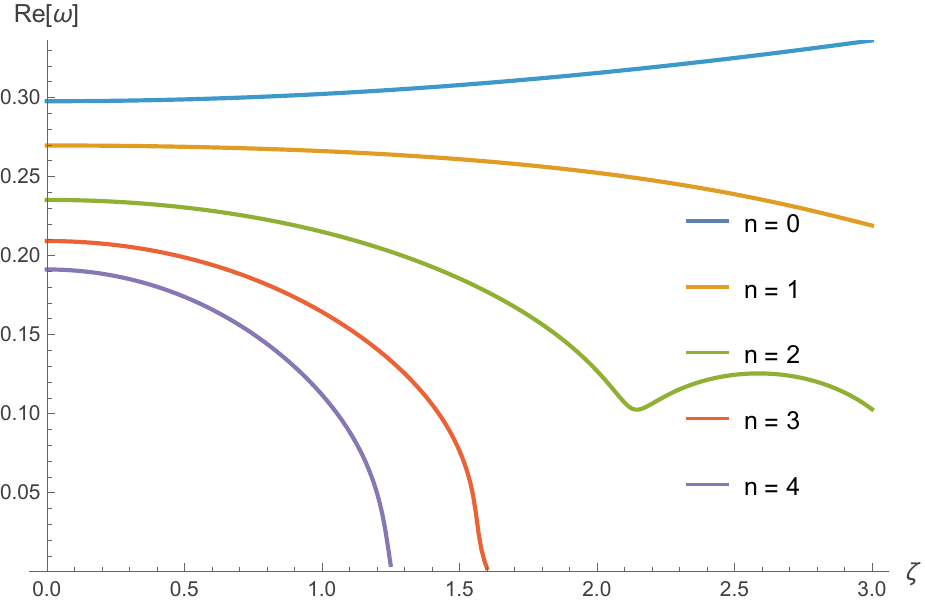}}
\hspace{0.02\linewidth}
\subfloat[]{\includegraphics[width=0.42\linewidth]{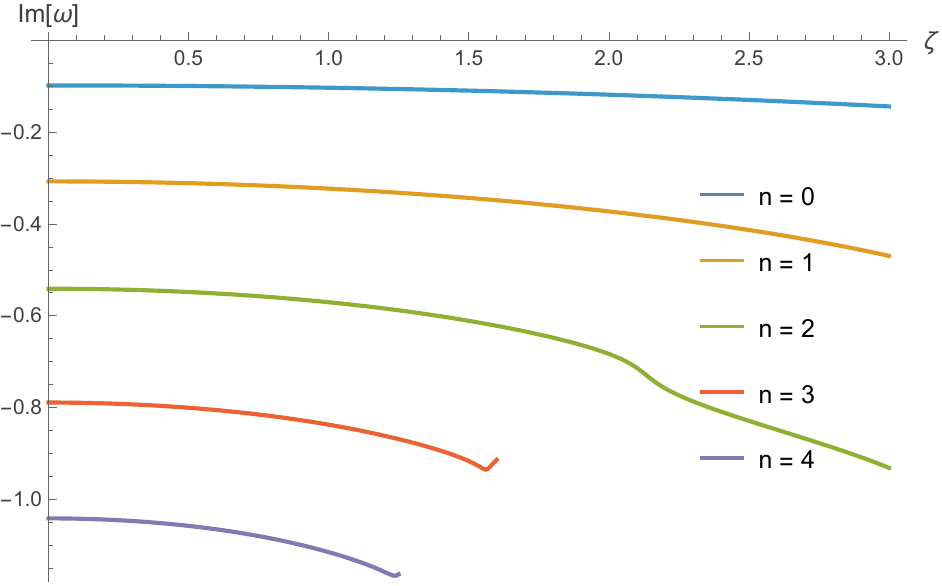}}
\vspace{4pt}
\subfloat[]{\includegraphics[width=0.42\linewidth]{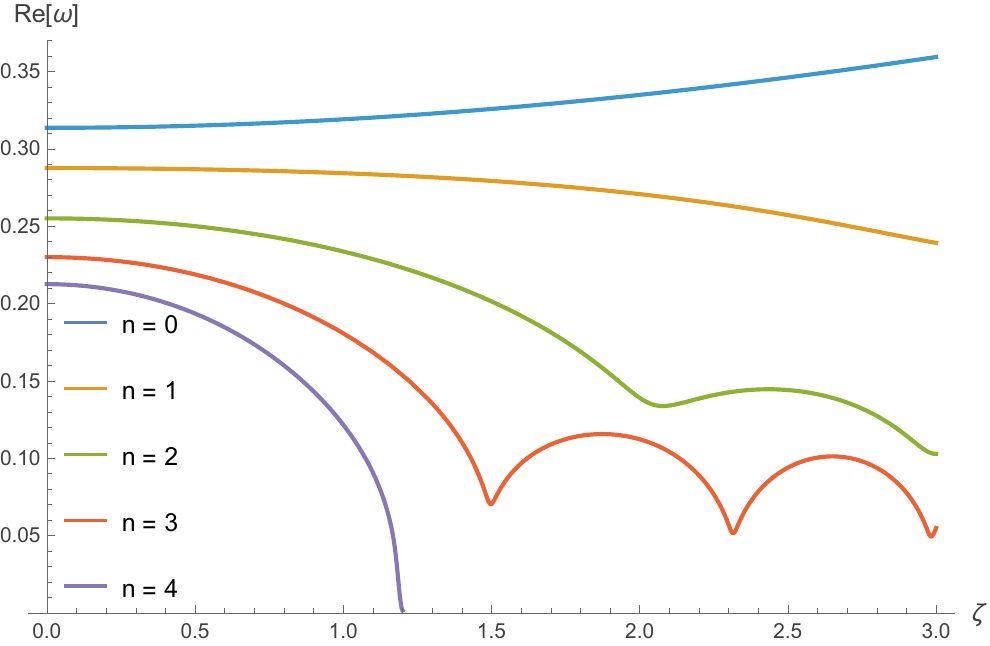}}
\hspace{0.02\linewidth}
\subfloat[]{\includegraphics[width=0.42\linewidth]{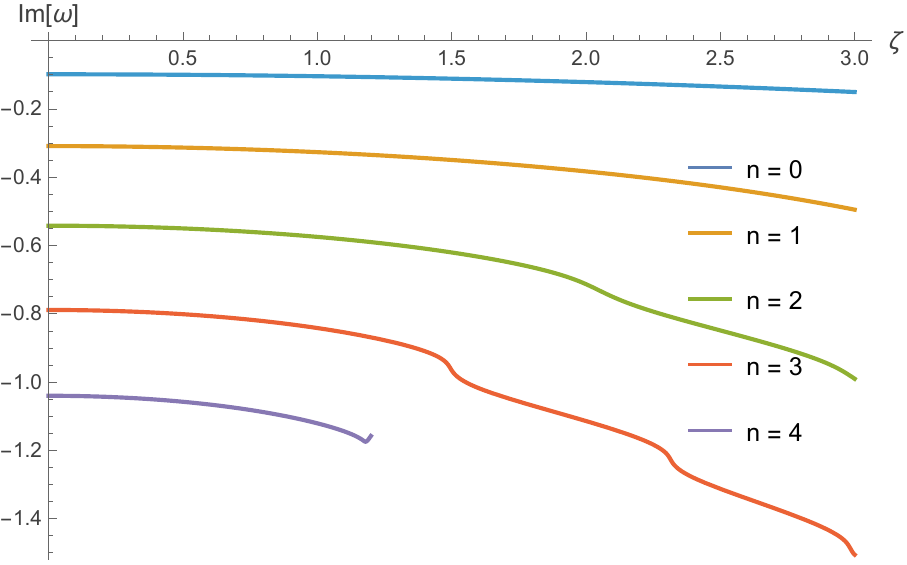}}
\vspace{4pt}
\subfloat[]{\includegraphics[width=0.42\linewidth]{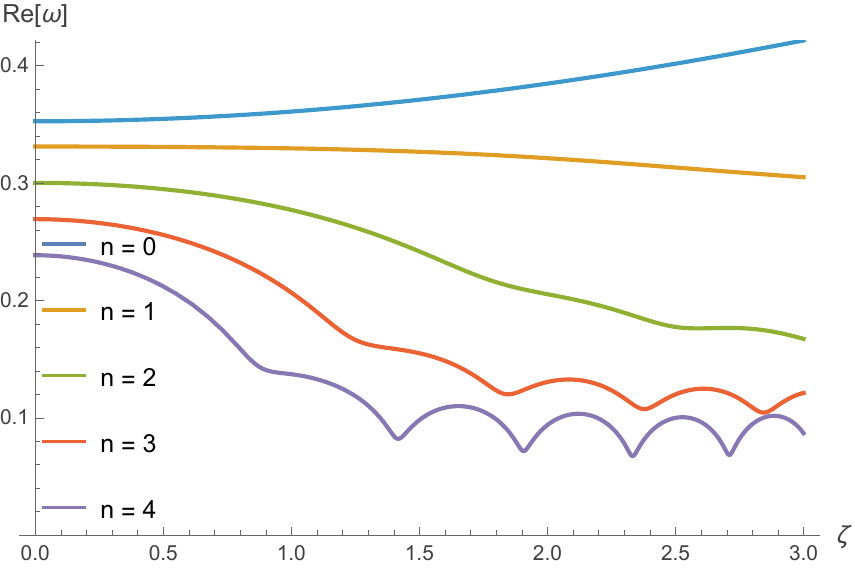}}
\hspace{0.02\linewidth}
\subfloat[]{\includegraphics[width=0.42\linewidth]{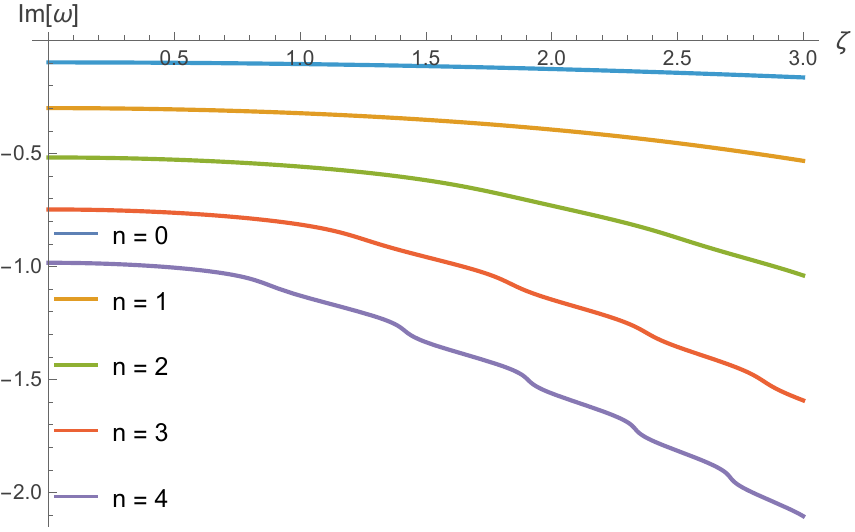}}
\caption{The QNMs of the fundamental mode and the first four overtones as a function of $\zeta$ for $l=1$. The charges in the first, second, third, and fourth rows are $0$, $0.3$, $0.6$, and $0.9$, respectively.}
\label{fig7}
\end{figure*}

Note also that for the $l=1$ modes, the real parts of the quasinormal frequecies of the first overtone and the second overtone cease to approach zero even when the charge is zero while for the $l=0$ modes, the real parts of the frequencies of all the overtones go to zero as $\zeta$ increases when the charge is zero. So the turn-on of the quantum parameter $\zeta$ makes the real part of the overtone quasinormal frequency have a tendency to approach zero while increasing multipole number $l$ tends to prohibit this tendency. This indicates that increasing the multipole number $l$ can suppress the quantum effects, which has been observed in several previous studies\cite{Fu2024,Gong2024,Zhang2024}. As we have pointed out, increasing the black hole charge $Q$ also prohibits this tendency, which indicates that increasing $Q$ can also suppress the quantum gravity effects. This suppression of quantum effects with the increase of $Q$ has also been observed in \cite{Zhu2025}, where the authors found that increasing $Q$ universally suppresses quantum-gravity-induced
spectral features—outbursts, non-monotonicity, and oscillations.
\subsection{\label{sect42}Neutral massive field}
\begin{figure*}[htb]
\centering
\subfloat[]{\includegraphics[width=0.42\linewidth]{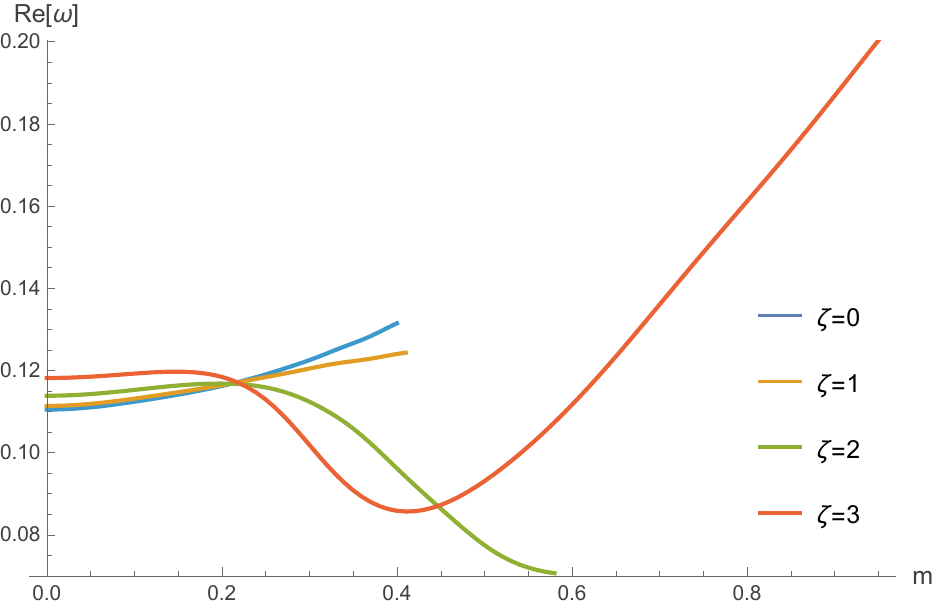}}
\hspace{0.02\linewidth}
\subfloat[]{\includegraphics[width=0.42\linewidth]{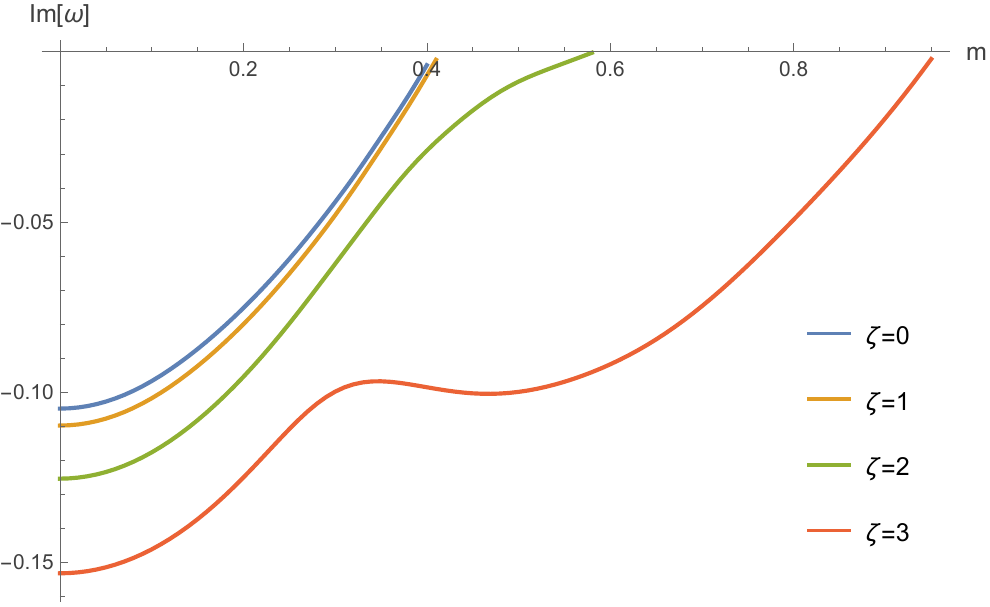}}
\vspace{4pt}
\subfloat[]{\includegraphics[width=0.42\linewidth]{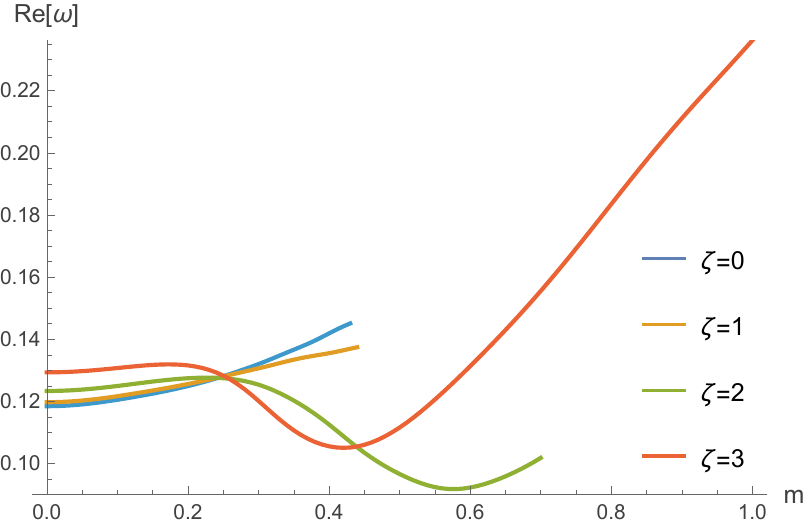}}
\hspace{0.02\linewidth}
\subfloat[]{\includegraphics[width=0.42\linewidth]{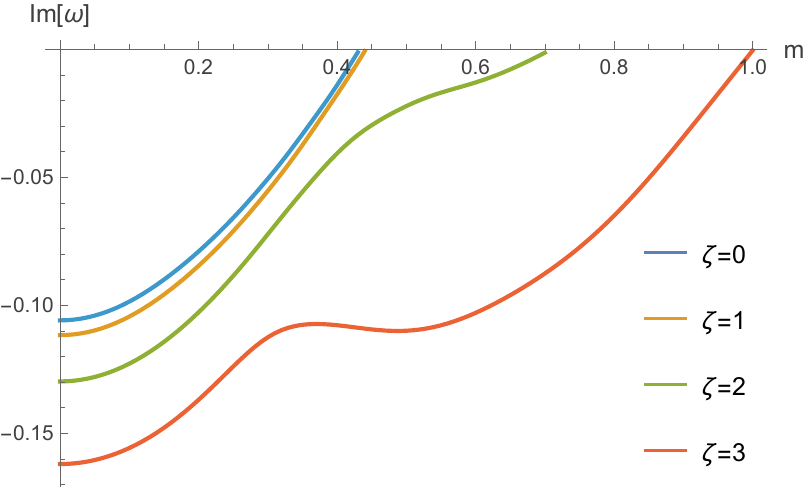}}
\vspace{4pt}
\subfloat[]{\includegraphics[width=0.42\linewidth]{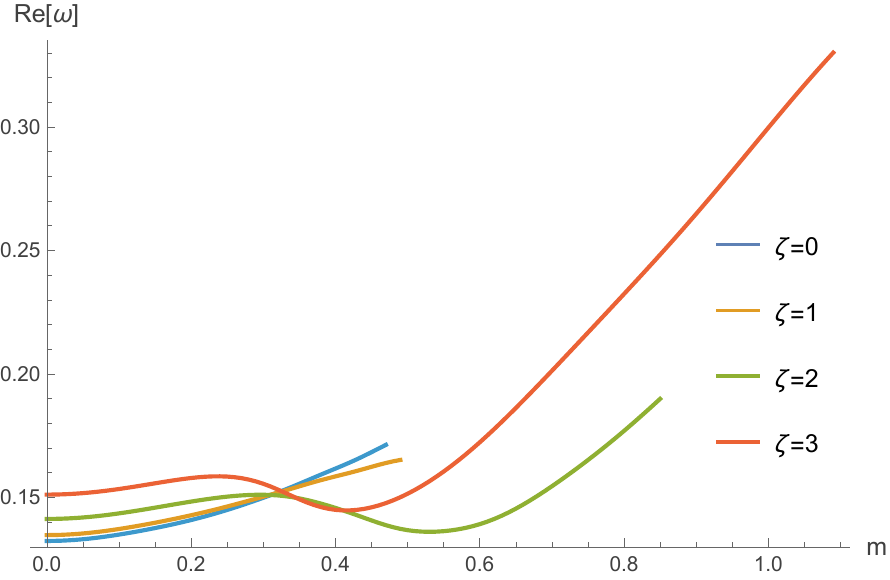}}
\hspace{0.02\linewidth}
\subfloat[]{\includegraphics[width=0.42\linewidth]{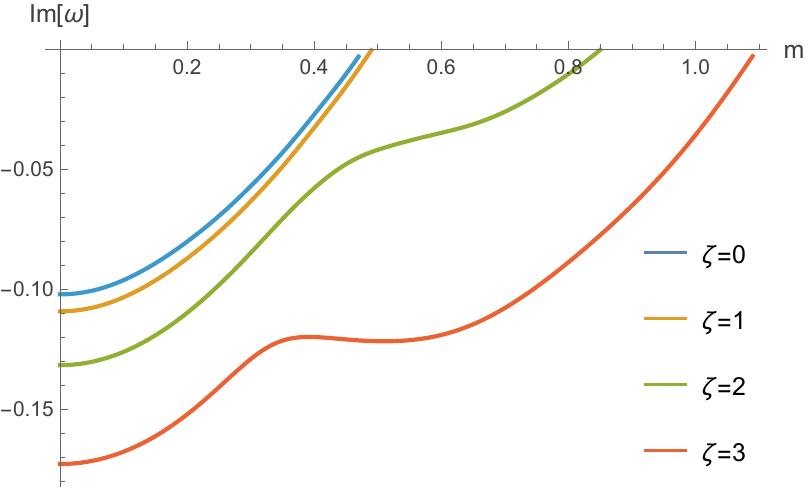}}
\vspace{4pt}
\subfloat[]{\includegraphics[width=0.42\linewidth]{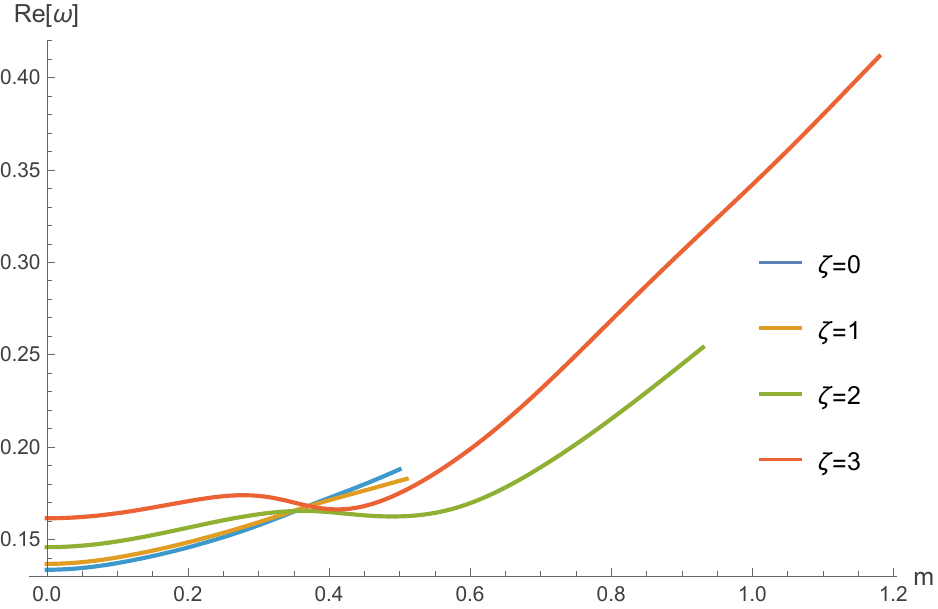}}
\hspace{0.02\linewidth}
\subfloat[]{\includegraphics[width=0.42\linewidth]{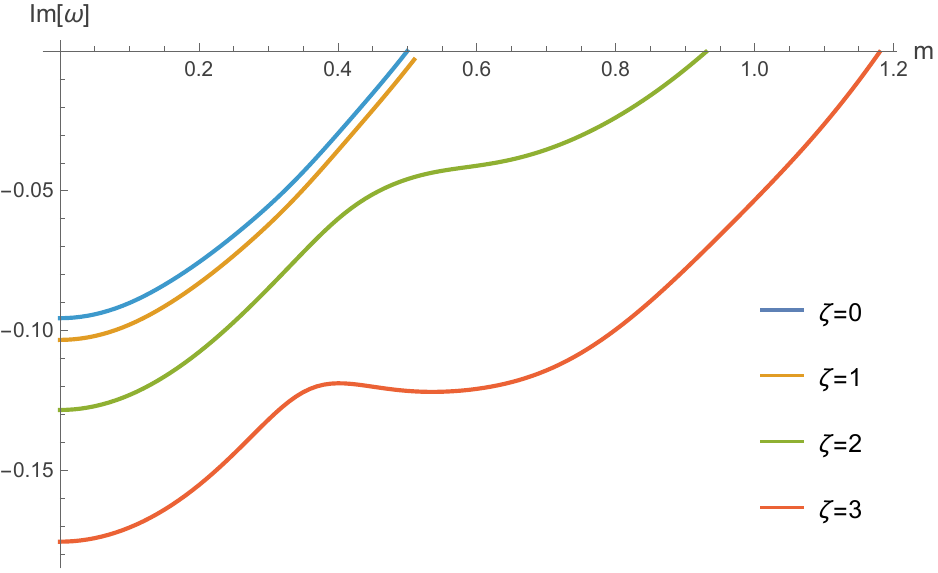}}
\caption{The QNMs of the fundamental mode of neutral massive field with $l=0$. The charges in the first, second, third, and fourth rows are $0$, $0.6$, $0.9$, and $0.99$, respectively.}
\label{fig8}
\end{figure*}

The QNMs as a function of the field mass $m$ are shown in Fig.\ref{fig8}. Regardless of the charge $Q$ of the black hole, for small quantum parameter $\zeta$, the real part of the fundamental mode frequency increases monotonically as the field mass increases while the imaginary part decreases monotonically and approaches zero at some critical values of $mass$. However, for large $\zeta$, both the real part and the imaginary part exhibit non-monotonic behavior. The real part first increases, then decreases, and then increases again as $m$ increases. The imaginary part first decreases, then increases and finally decreases to zero at some critical value as the field mass $m$ increases. As far as we know, this is the first time we observe non-monotonicity of QNMs as a function of field mass of scalar field in black holes of both general relativity and modified theories of gravity, although non-monotonicity is not rare in overtone modes\cite{Bolokhov2025,Konoplya2005,Zinhailo2024b}. We hereby have found that large quantum parameter $\zeta$ can induce non-monotonicity behavior of the QNMs as a function of $m$ in the fundamental mode. The QNMs of massive scalar field of the corresponding uncharged covariant quantum effective black hole has been examined in \cite{Bolokhov2025}. However, the author missed to notice this non-monotonicity for he did not consider the case of a large $\zeta$. 
\par When the imaginary part of the quasinormal frequency approaches zero, the QNMs oscillate without damping, leading to the pheonomenon called quasi-resonance. We thus verify that quasi-resonances still exist when both the black hole charge $Q$ and quantum parameter are turned on. From Fig.\ref{fig8}, we can see that for fixed charge $Q$, as $\zeta$ increases, the critical mass value of $m$ at which the imaginary part approaches zero increases. If we denote the critical value as $m_c$, we also note that for the uncharged black hole case, $dm_c/d\zeta$ increases monotonically as $\zeta$ increases, while for the charged case, $dm_c/d\zeta$ first increases and then decreases. For fixed $\zeta$, $m_c$ also increases with increasing $Q$. Thus, both $Q$ and $\zeta$ tend to increase $m_c$. We also note that increasing the charge $Q$ does not alter the qualitative behavior of the QNMs as a function of $m$, although for $\zeta=3$ and $Q=0$, the real part does not exhibit non-monotonicity as the charged cases do, which may be due to the possibility that the imaginary goes to zero before the non-monotonicity of the real part emerges. 
\subsection{\label{sec43}Charged massless field}
\begin{figure*}[htb]
\centering
\subfloat[]{\includegraphics[width=0.38\linewidth]{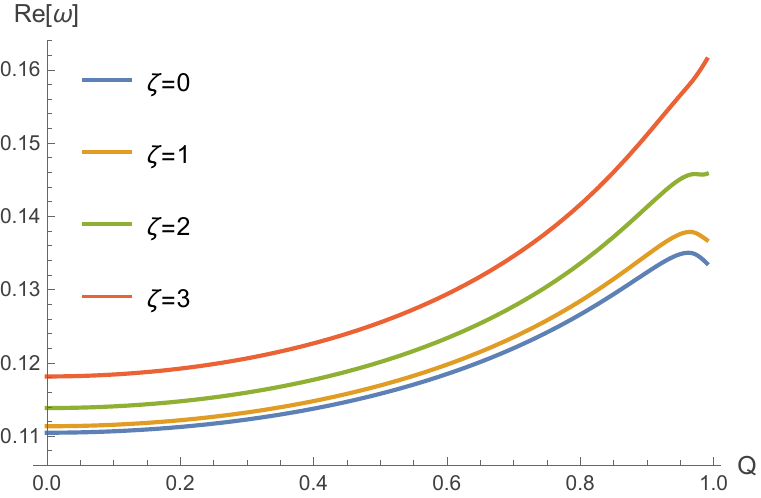}}
\hspace{0.02\linewidth}
\subfloat[]{\includegraphics[width=0.38\linewidth]{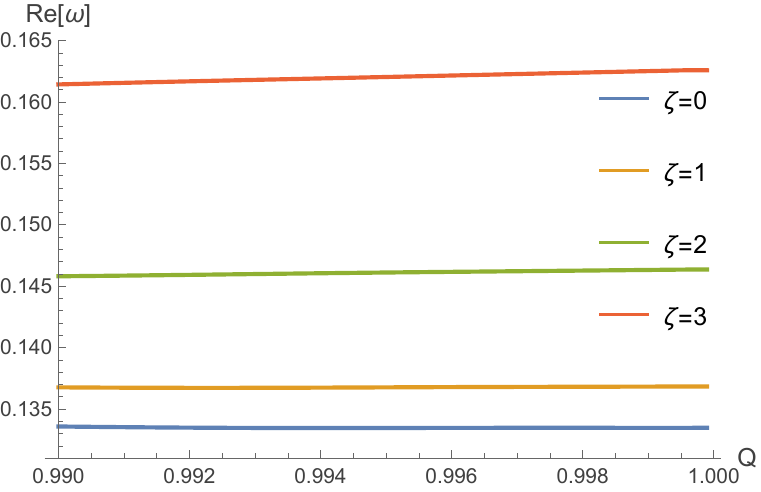}}
\vspace{2pt}
\subfloat[]{\includegraphics[width=0.38\linewidth]{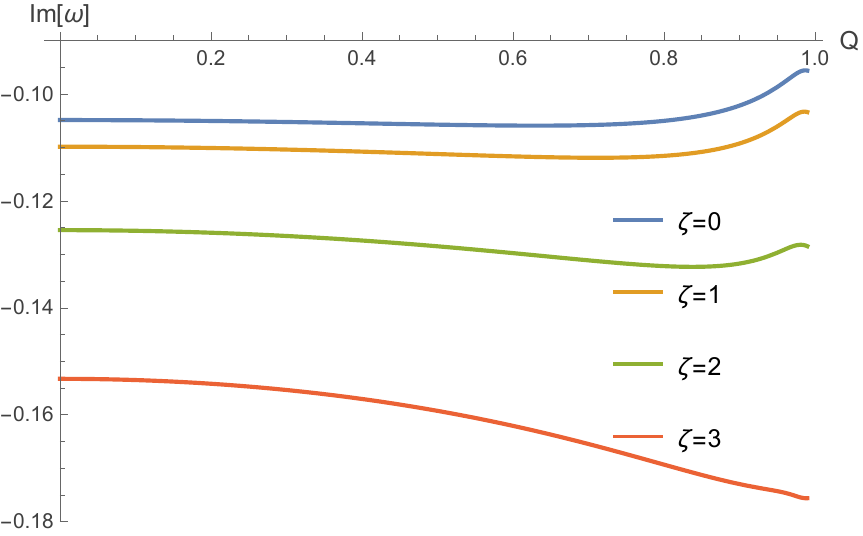}}
\hspace{0.02\linewidth}
\subfloat[]{\includegraphics[width=0.38\linewidth]{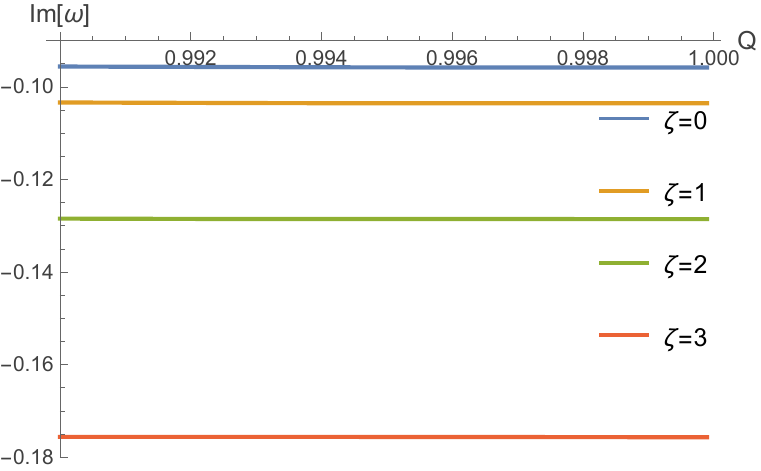}}
\caption{The QNMs of uncharged massless scalar field as a function of black hole charge $Q$ with $l=0$. In the left panel, the range of $Q$ is $[0,0.99]$. In the right panel, the range of $Q$ is $[0.99,0.9999]$}
\label{fig9}
\end{figure*}

\vspace{-8pt}

\begin{figure*}[htb]
\centering
\subfloat[]{\includegraphics[width=0.38\linewidth]{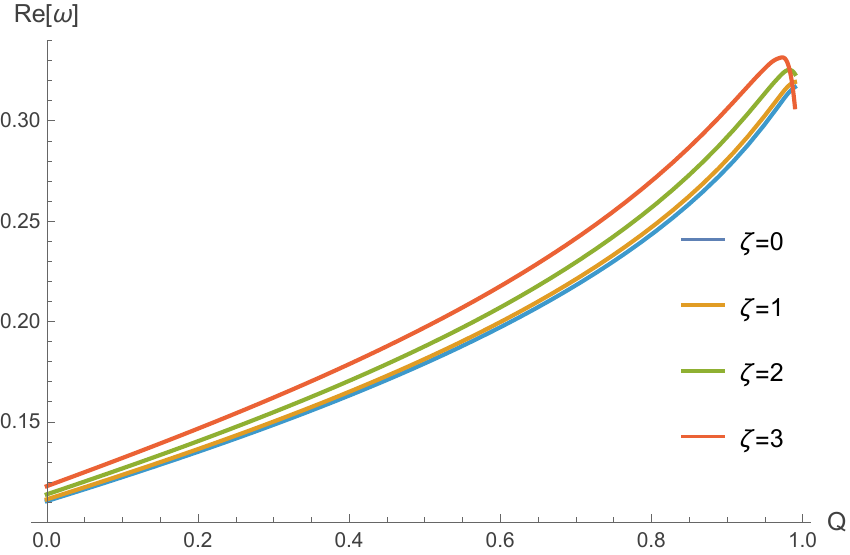}}
\hspace{0.02\linewidth}
\subfloat[]{\includegraphics[width=0.38\linewidth]{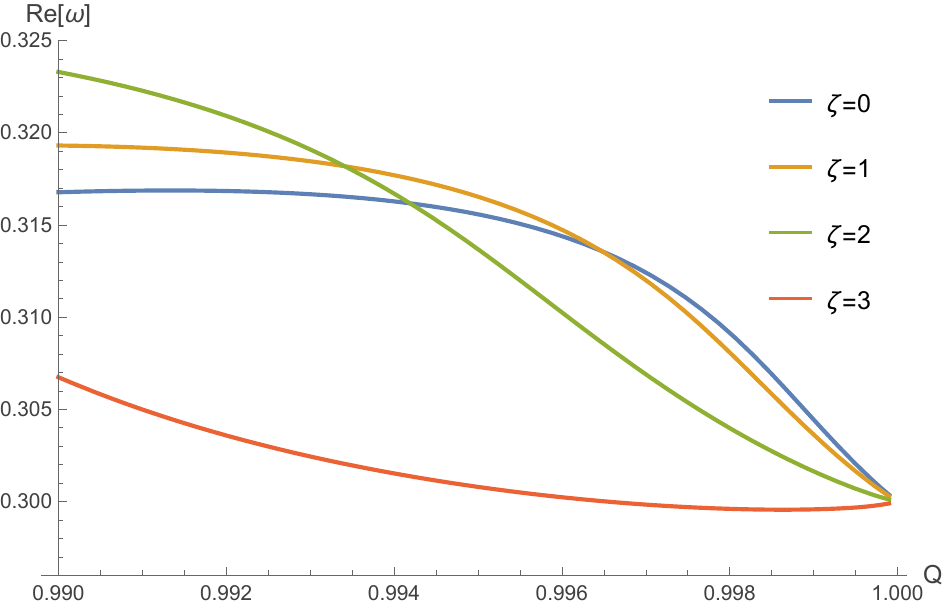}}
\vspace{2pt}
\subfloat[]{\includegraphics[width=0.38\linewidth]{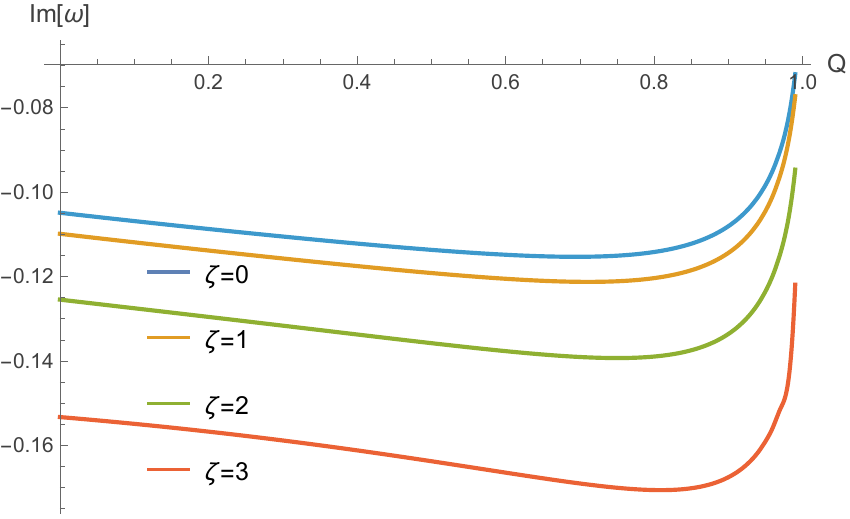}}
\hspace{0.02\linewidth}
\subfloat[]{\includegraphics[width=0.38\linewidth]{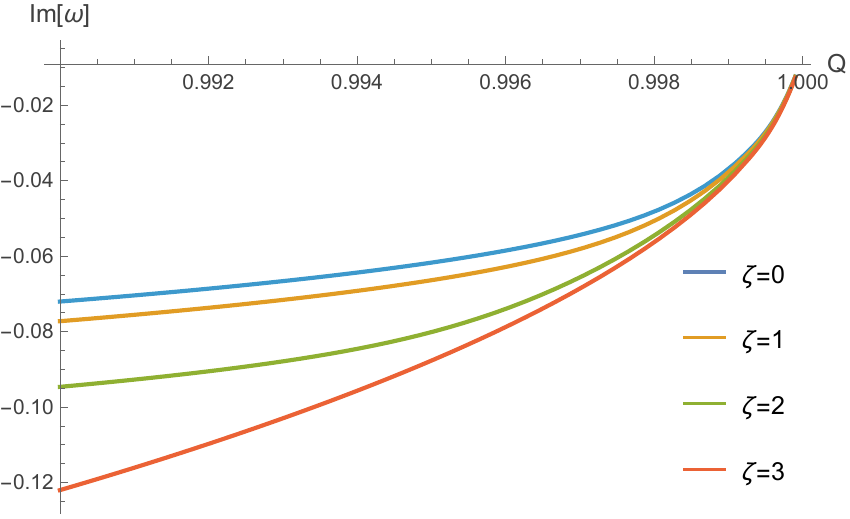}}
\caption{The QNMs of charged massless scalar field with field charge $e=0.3$ as a function of black hole charge $Q$ with $l=0$. In the left panel, the range of $Q$ is $[0,0.99]$. In the right panel, the range of $Q$ is $[0.99,0.9999]$}
\label{fig10}
\end{figure*}

\begin{figure*}[htb]
\centering
\subfloat[]{\includegraphics[width=0.42\linewidth]{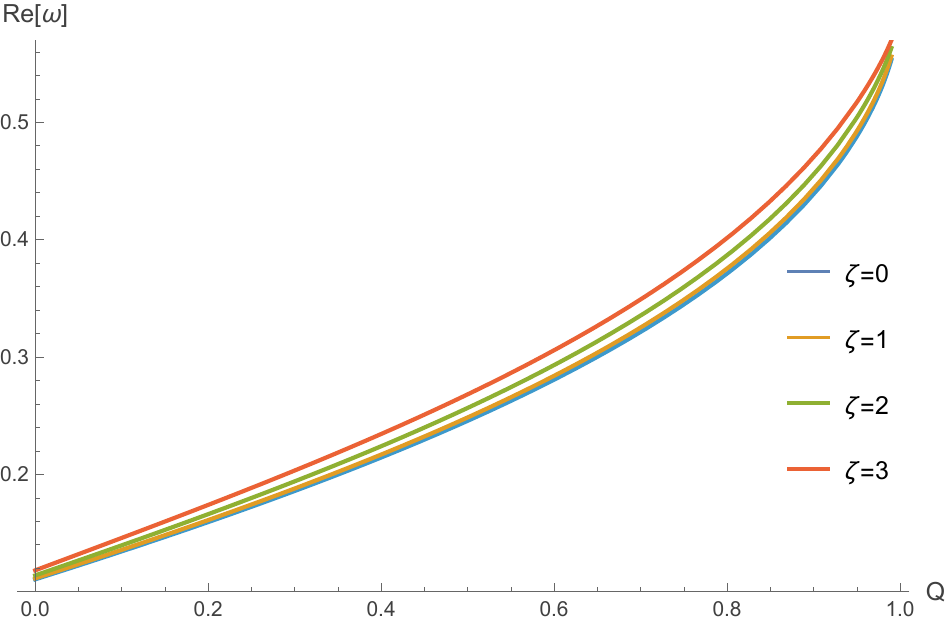}}
\hspace{0.02\linewidth}
\subfloat[]{\includegraphics[width=0.42\linewidth]{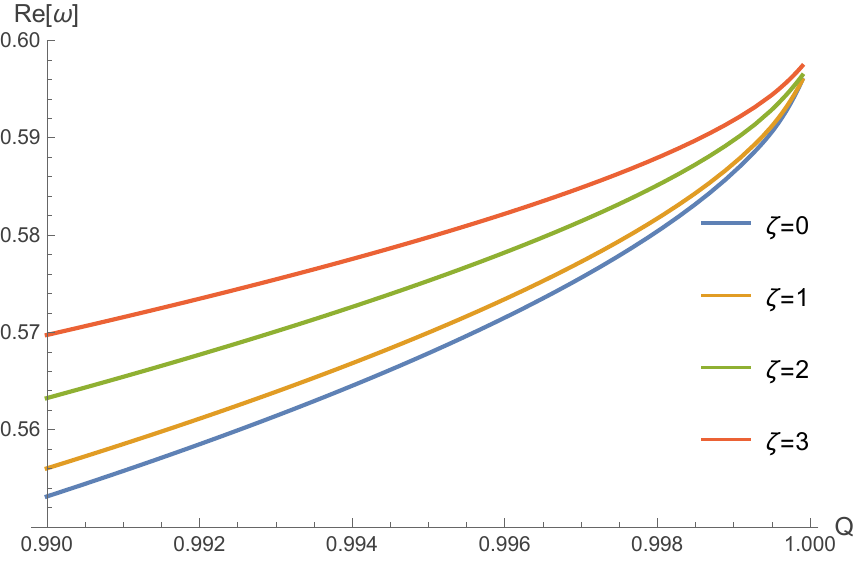}}
\vspace{4pt}
\subfloat[]{\includegraphics[width=0.42\linewidth]{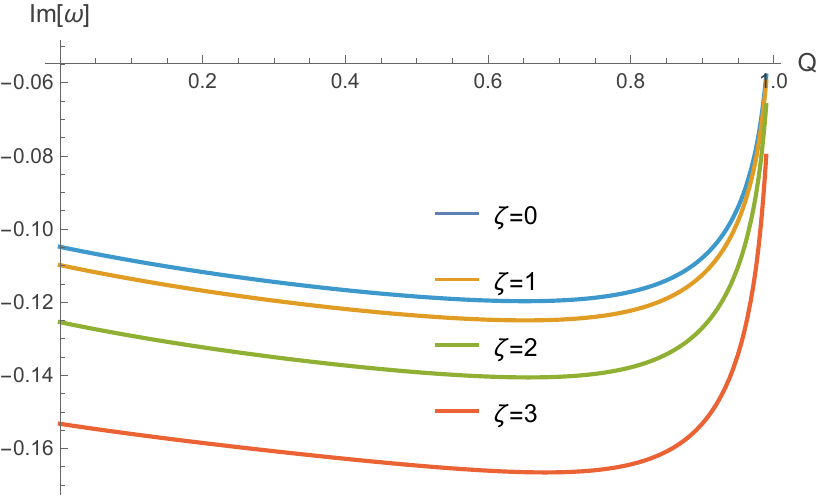}}
\hspace{0.02\linewidth}
\subfloat[]{\includegraphics[width=0.42\linewidth]{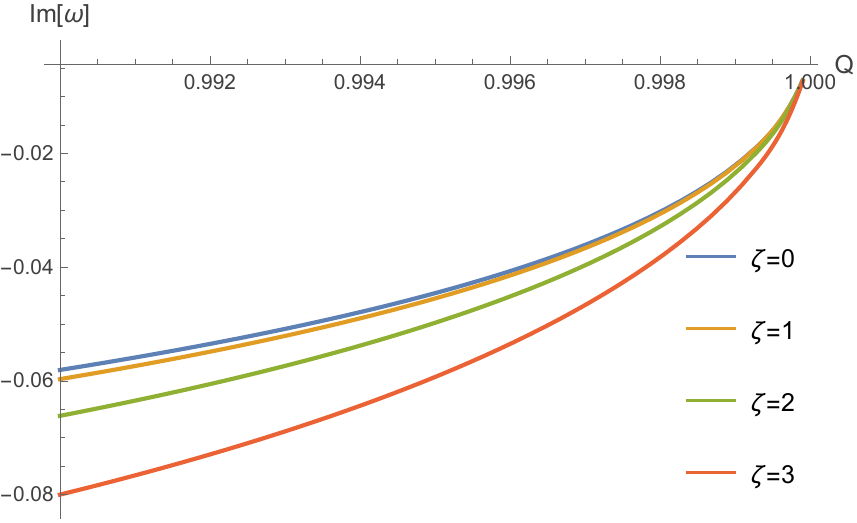}}
\caption{The QNMs of charged massless scalar field with field charge $e=0.6$ as a function of black hole charge $Q$ with $l=0$. In the left panel, the range of $Q$ is $[0,0.99]$. In the right panel, the range of $Q$ is $[0.99,0.9999]$}
\label{fig11}
\end{figure*}

Fig.\ref{fig9} shows the QNMs of neutral massless scalar field as a function of black hole charge $Q$ with $l=0$, while Fig.\ref{fig10} and Fig.\ref{fig11} show the QNMs of charged massless scalar field with field charge $e=0.3$ and $e=0.6$ ,respectively. We can clearly see that as $Q$ approaches 1, QNMs of charged and uncharged scalar field have different asymptotic behaviors. As the maximal charge is approached, the real part of QNMs of charged scalar field approaches $eQ/R_+
\approx
e$, while the imaginary part approaches zero. However, the real part of QNMs of uncharged scalar field does not approaches zero as the maximal charge is approached. The quasinormal frequencies at near-extremal black hole charge $Q=0.9999$ are listed in Table~\ref{tab1}. This indicates that there exists some modes that are solutions to the continued fraction equation when the field charge is not zero and that these modes does not exist when the field charge is zero. These strange modes was first identified for RN black hole in \cite{Konoplya2013} and later confirmed to exist in the Dirac field case too in \cite{Richartz2014}. In \cite{Richartz2014}, the authors verified that these modes are not numerical errors of the continued fraction method, but indeed solutions of the equation of motion analytically. We hereby verify that these modes still exist even when the quantum parameter $\zeta$ is turned on. As pointed out in \cite{Richartz2014}, as the imaginary of these modes approaches zero as $Q$ approaches 1, this indicates that exactly extremal black holes may have vanishing imaginary part, leading to the quasi-resonance phenomenon, similar to the case of massive scalar field.

\par
\begin{figure*}[htb]
\centering
\subfloat[]{%
  \includegraphics[width=0.42\textwidth]{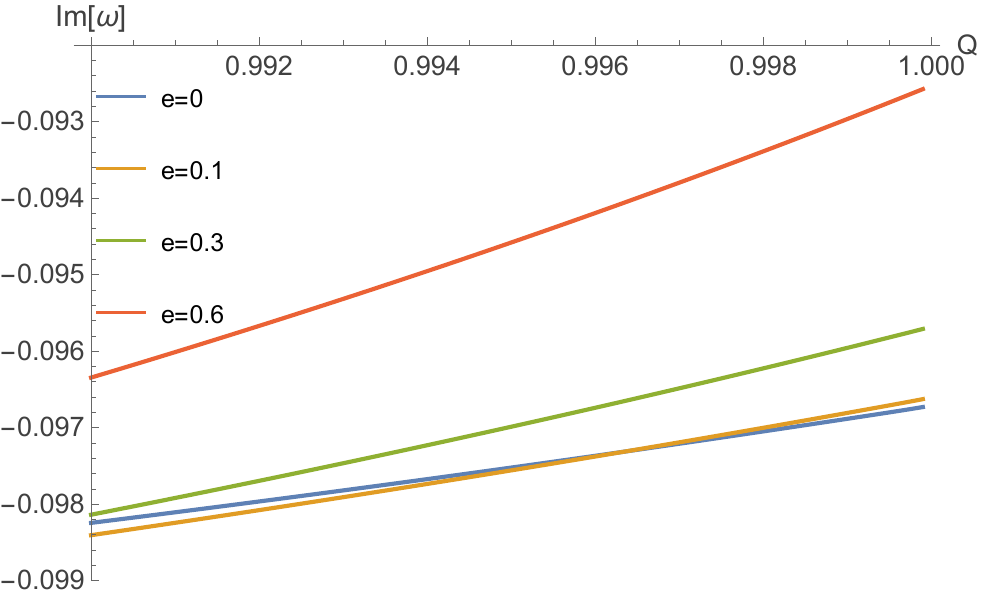}
  \label{fig12a}
}
\hspace{0.02\textwidth}
\subfloat[]{%
  \includegraphics[width=0.42\textwidth]{fig121.pdf}
  \label{fig12b}
}
\caption{The asymptotic behavior of the imaginary part of the fundamental mode with $l=3$ as $Q$ approaches the extremal value. In the left panel, the quantum parameter $\zeta=1$. In the right panel, the quantum parameter $\zeta=0$. $Q$ runs from $0.99$ to $0.9999$}
\label{fig12}
\end{figure*}

 In~\cite{Konoplya2002decay}, the author found that for RN black holes, the imaginary part of the fundamental mode of charged scalar field approaches that of neutral scalar field in the limit of extremal black hole, no matter what the field charge is, with the use of WKB method. So we want to check if this phenomenon still occur with the quantum parameter turned on. Fig.\ref{fig12a} shows the asymptotical behaviors of the imaginary part of the fundamental mode with $l=3$ for charged and neutral field at fixed $\zeta$. 
\begin{figure*}[htb]
\centering
\subfloat[]{
  \includegraphics[width=0.31\textwidth]{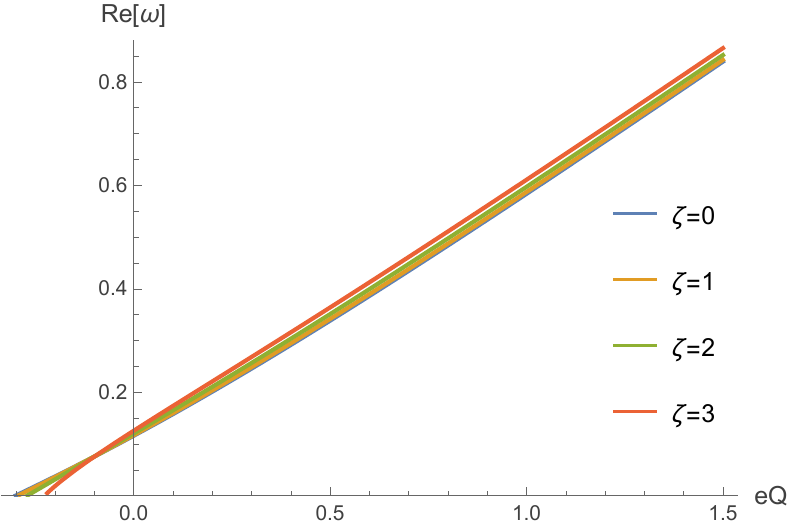}
  \label{fig13a}
}
\hfill
\subfloat[]{
  \includegraphics[width=0.31\textwidth]{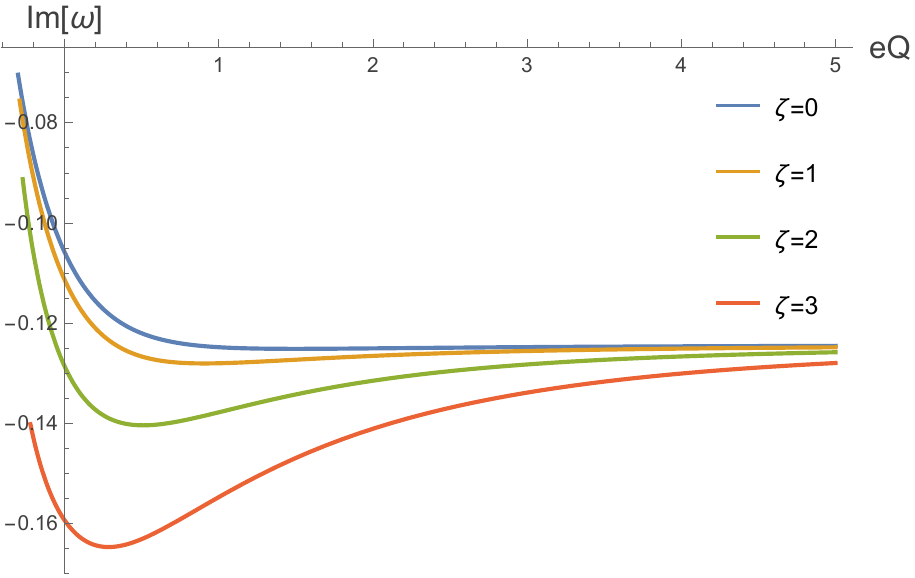}
  \label{fig13b}
}
\hfill
\subfloat[]{
  \includegraphics[width=0.31\textwidth]{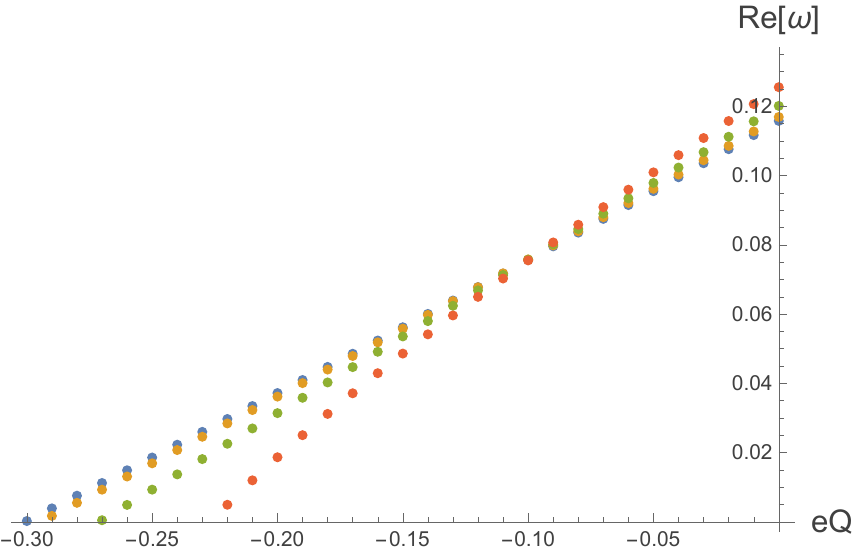}
  \label{fig13c}
}
\caption{The QMNs of the fundamental mode of charged scalar field with $l=1$ as a function of $eQ$. The right panel is the enlargement of the region where the real part approaches zero}
\label{fig13}
\end{figure*}
\begin{figure*}[htb]
\centering
\subfloat[]{%
  \includegraphics[width=0.42\textwidth]{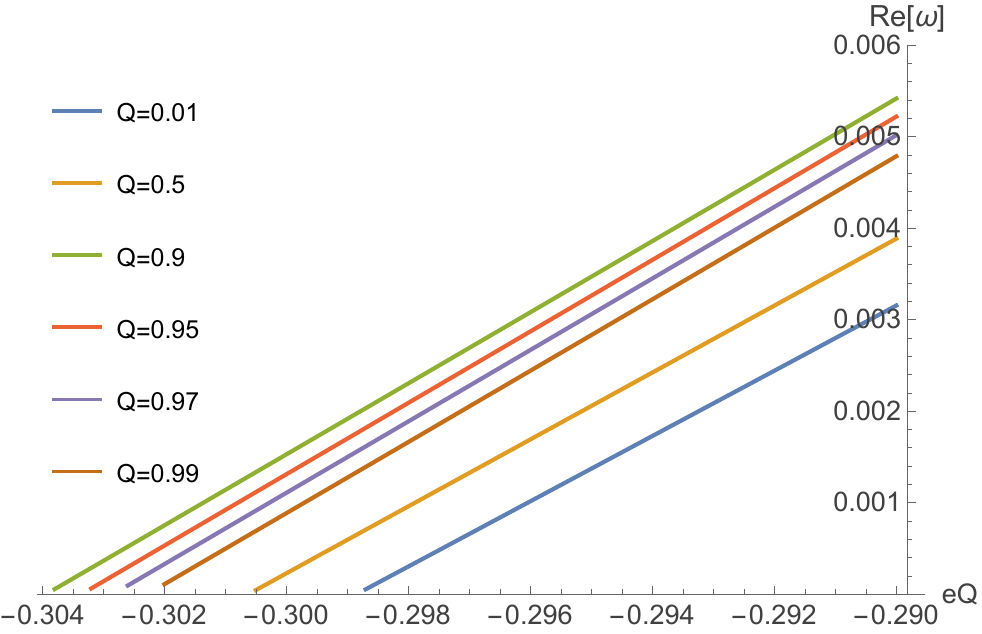}
  \label{fig14a}
}
\hspace{0.02\textwidth}
\subfloat[]{%
  \includegraphics[width=0.42\textwidth]{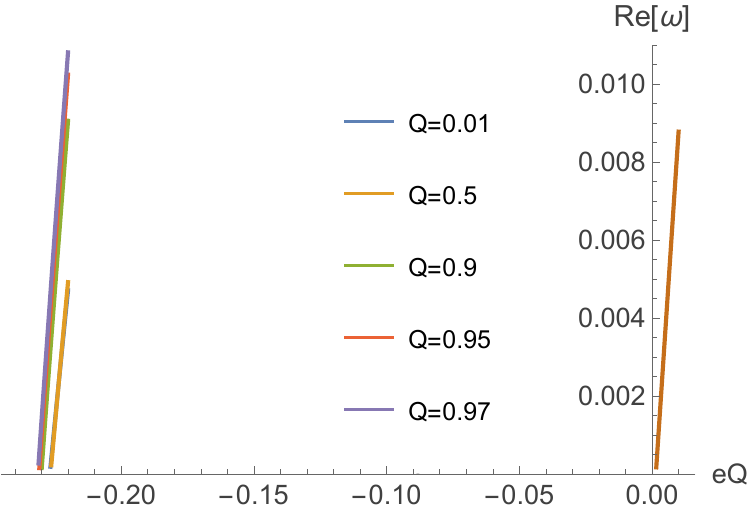}
  \label{fig14b}
}
\caption{The real part of the fundamental mode with $l=0$ as a function of $eQ$ in the region where the real part approaches zero. In the left panel, $\zeta=0$. In the right panel, $\zeta=3$.}
\label{fig14}
\end{figure*}

We can clearly see that the imaginary part of the charged field does not approach that of the neutral field, contradicting with what was found in \cite{Konoplya2002decay}. Although it may be due to the introduction of the quantum parameter, there is still a possibility that this is still true for RN black hole. So we further check if the imaginary part of the charged field really approaches that of the neutral field in the limit of extremal black hole in the RN black hole case. The result is shown in Fig.\ref{fig12b}. Clearly, the imaginary part of charged and uncharged field does not have the same asymptotic value as the extremal charge is approached. Because the continued fraction method we use in this work is more reliable than the third order WKB method used in \cite{Konoplya2002decay}. We tend to believe that our result is the correct one. And the wrong conclusion in \cite{Konoplya2002decay} is due to the inaccuracy of the third order WKB method. For example, in \cite{Konoplya2002decay}, the imaginary part at $e=0$, $l=3$ and $Q=0.995$ is $ 0.08938$, and the imaginary part at $e=0.3$, $l=3$ and $Q=0.995$ is $0.08943$, while the values we obtain from the continued fraction method are $0.08938$ and $0.08893$, respectively. The near extremal quasinormal frequencies at different field charge $e$ for both $\zeta=0$ and $\zeta=1$ are listed in Table~\ref{table2}.

\par It was first noted in \cite{Konoplya2013} that for charged scalar field in RN black hole, the real part of the fundamental mode approaches zero and then disappears from the spectrum at some critical value of $eQ$, then in \cite{Richartz2014}, the author verified that this phenomenon also exists in the Dirac field case. Hence, we are curious about the existence of this phenomenon when the RN black hole is quantum-corrected. Fig.\ref{fig13} shows the QNMs of the fundamental mode with $l=0$. We can see that this phenomenon still persists with the quantum parameter $\zeta$ turned on. 

For the RN black hole case, we find that the critical value is around $-0.3$, which is in accordance with the result in \cite{Richartz2014}. And the critical values for $\zeta=1$, $\zeta=2$ and $\zeta=3$ are around $-0.29$, $-0.27$ and $-0.22$. The quantum parameter $\zeta$ modifies the critical value. If we denote the critical value as $\alpha_c$, it seems that $d\alpha_c/d\zeta$ increases as $\zeta$ increases. In \cite{Richartz2014}, the authors also state that as $Q$ varies, the critical values is almost unchanged. However, we find that this is not turn when we include quantum correction. The most striking feature we can see in Fig.\ref{fig14} is that for near extremal charge $Q=0.99$, the critical value is significantly different from those of lower charges. In this near extremal case, the fundamental mode disappears at a positive value of $eQ$, while for all other cases, the disappearances occur at a negative value of $eQ$. Different from the RN black hole case that the critical value is almost unaffected by the charge $Q$, this seems to indicate that the quantum parameter can enhance the effect induced by increasing $Q$ and that with the quantum parameter turned on, the behavior of the QNMs as a function of $eQ$ at near extremal charge can be qualitatively different from that of the case at non-extremal charge.
\section{\label{sec5}Conclusions}
In this work, we have studied the QNMs of neutral massless scalar field, neutral massive scalar field and charged massless scalar field for a charged covariant quantum-corrected black hole using the continued fraction method with Nollert improvement. For the neutral massless scalar field we find that for smaller charges, the behavior of QNMs as a function of the Quantum parameter $\zeta$ is qualitatively the same as that in the case of an uncharged black hole. The real parts of the overtones still approach zero with the black hole charge $Q$ turned on as in the case of an uncharged black hole. However, for larger charges, the behavior becomes qualitatively different from that in the uncharged black hole case. The real part of some overtones cease to approach zero in the whole range of $\zeta$ under consideration. Increasing the multipole number can also prohibit the real parts of the overtones from approaching zero. We hereby find that both the black hole charge and increasing the multipole number $l$ can suppress the effect induced by $\zeta$.\par
For the neutral massive field, we find that with both $Q$ and the quantum parameter $\zeta$ turned on, quasi-resonances still emerge at some critical value of field mass $m$. And the quantum parameter $\zeta$ can significantly alter the critical $m$ at which quasi-resonances emerge. And above all, for the first time, we observe non-monotonic behaviors of both real and imaginary part of the QNMs of the fundamental mode as a function of field mass $m$ when $\zeta$ is sufficiently large.
For the charged massless field, we confirmed the existence of some strange modes that are solutions of the continued fraction equation in the case of charged scalar field, but does not exist in the neutral scalar field case. The real parts of these modes approach $eQ/R_+
\approx
e$ as the extremal charge is approached, while the imaginary parts approaches zero, which may hint the existence of quasi-resonances in the exactly extremal black hole case. Also by accident, we find that the statement in \cite{Konoplya2002decay} that the imaginary part of the fundamental mode of charged scalar field approaches that of neutral scalar field is probably wrong due to the limitation of the third order WKB method. We also confirmed that with the introduction of quantum correction, the real part of the fundamental mode still approaches zero at some critical value of $eQ$ and the quantum parameter $\zeta$ can change the critical $eQ$. We also find that for fixed $\zeta$, the critical value for near extremal charge $Q$ can be very different from those for non-extremal charges. Because in the RN black hole case, the charge have negligible effect on the critical value, we tend to conceive that the turn-on of $\zeta$ can enhance the effect induced by $Q$ on the QNMs of charged scalar field.\par
Future directions may include the investigations of perturbations of Dirac field, electromagnetic field and gravitational field for this charged covariance quantum-corrected black hole. 
\begin{acknowledgments}
This work is supported by the starting fund for young talents of Huanggang Normal University.
\end{acknowledgments}

\clearpage

\nocite{*}
\bibliography{apssamp}

\end{document}